\documentclass{article}

\usepackage{arxiv}

\usepackage[utf8]{inputenc} 
\usepackage[T1]{fontenc}    
\usepackage{hyperref}       
\usepackage{url}            
\usepackage{booktabs}       
\usepackage{amsfonts}       
\usepackage{amsmath,amssymb}
\usepackage{nicefrac}       
\usepackage{microtype}      
\usepackage{cleveref}       
\usepackage{lipsum}         
\usepackage{graphicx}
\usepackage{doi}

\usepackage{latexsym}
\usepackage{multicol,multirow}
\usepackage{mathrsfs}
\usepackage{amsthm}
\usepackage{apacite}
\usepackage[authoryear]{natbib}
\usepackage{rotating}
\usepackage{appendix}
\usepackage{ifpdf}
\usepackage{newtxmath}
\usepackage{textcomp}%
\usepackage{xcolor}%

\usepackage{epstopdf}
\usepackage{bm}

\def\clap#1{\hbox to 0pt{\hss#1\hss}}

\providecommand{\mat}[1]{\bm{#1}}%

\providecommand{\mC}{\ensuremath{\mat{C}}}

\newcommand*\circled[1]{\tikz[baseline=(char.base)]{
		\node[shape=circle,draw,inner sep=2pt] (char) {#1};}}

\usepackage{savesym}
\savesymbol{medsize}
\usepackage{nccmath}
\usepackage{subfigmat}
\usepackage{tikz}
\usepackage{placeins}	
\usepackage{booktabs}
\usepackage[export]{adjustbox}
\let\oldcaption=\caption
\renewcommand{\caption}[1]{\oldcaption[#1]{#1.}}

\title{Density reconstruction from schlieren images through Bayesian nonparametric models}

\date{}

\author{ \href{https://orcid.org/0000-0002-0206-7140 }{\includegraphics[scale=0.06]{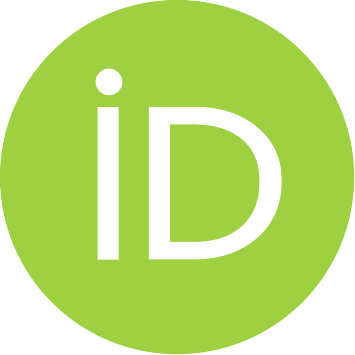}\hspace{1mm}Bryn Noel Ubald}\thanks{https://www.brynnoelubald.com/} \\
	Data-Centric Engineering\\
	The Alan Turing Institute\\
	96 Euston Road, London, NW1 2DB \\
	\texttt{bubald@turing.ac.uk} \\
	\And
	\href{https://orcid.org/0000-0002-7351-012X}{\includegraphics[scale=0.06]{orcid.pdf}\hspace{1mm}Pranay Seshadri} \\
	Department of Mathematics\\
	Imperial College London\\
	Exhibition Rd, London, SW7 2BX \\
	\texttt{p.seshadri@imperial.ac.uk} \\
	\And
	Andrew Duncan \\
	Department of Mathematics\\
	Imperial College London\\
	Exhibition Rd, London, SW7 2BX \\
	\texttt{a.duncan@imperial.ac.uk} \\
}

\renewcommand{\headeright}{}

\hypersetup{
pdftitle={Density reconstruction from schlieren images through Bayesian nonparametric models},
pdfsubject={physics.flu-dyn, cs.CV},
pdfauthor={Bryn N. Ubald, Pranay Seshadri, Andrew Duncan},
pdfkeywords={Gaussian Processes, Schlieren, Bayesian inference, Kriging, Gradient Enhanced Kriging, GEK},
}

\begin{document}
\maketitle

\begin{abstract}
    This study proposes a radically alternate approach for extracting quantitative information from schlieren images. The method uses a scaled, derivative enhanced Gaussian process model to obtain true density estimates from two corresponding schlieren images with the knife-edge at horizontal and vertical orientations. We illustrate our approach on schlieren images taken from a wind tunnel sting model, a supersonic aircraft in flight, and a high-order numerical shock tube simulation.
\end{abstract}


\section{Introduction} \label{sec:intro}
Since its inception in the late-17th century by Robert Hooke, schlieren has been used in the sciences, engineering and even the arts. The types of flows captured have been diverse spanning the shock waves formed around a supersonic aircraft; the mixing of turbulent jets; plume formation over a candle, and even the dissolution of particles in a solution. Scientifically, these images have shaped our understanding of fluid mechanics. More broadly, they have, and continue to, inspire and captivate a collective curiosity about the natural world.

A schlieren is governed by refractive behaviour of light when passing through a medium. For instance, when encountering a homogeneous medium, light rays will pass through uniformly, but if the light rays pass through regions of inhomogeneity, they will refract in proportion to the gradient of the refractive index of the inhomogeneities \citep{Rienitz1975, Settles2001}.

\begin{figure}[!ht]
	\begin{center}
		\includegraphics[width=0.75\columnwidth, scale=0.30]{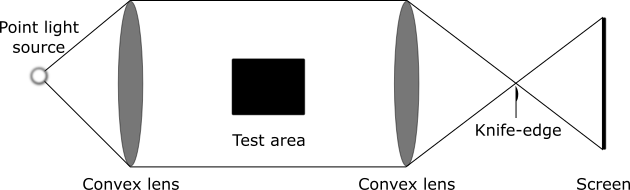}
		\caption{Schematic showing a simple schlieren experimental setup}
		\label{fig:setup}
	\end{center}
\end{figure}

Flow phenomena such as shock waves, jets and plumes can cause variations in the density of air, leading to refractions that can be captured by a focused, one-to-one optical image. Consider the simplest of schlieren setups: a point light source-based experiment that requires two convex lenses, a knife-edge, and a camera, as shown in Figure~\ref{fig:setup}. With the point light-source being the initial \emph{illuminator}, the light is collimated by the first lens, and then refocused to a single point by the second lens where a knife-edge is positioned. The light rays then travel past the knife-edge to cast an inverted image upon an appropriately placed screen. The area between the two lenses represents the test section where the object of interest should be placed. The knife-edge, which can be a standard razor, misses the upward deflected array, but blocks the downward deflected array. As a result the upward deflected array appears on the screen as a bright spot, while the downward deflected array leaves a dark spot. The knife-edge effectively transforms an invisible phase difference between the light rays to a visible amplitude change, which is recorded on the screen as a change in the refractive index\footnote{The refractive index is given by the speed of light through a given medium divided by the speed of light in vacuum}. Depending on the orientation of the knife-edge, either the horizontal or the vertical gradient of the refractive index can be visible. Although this description is of the most primitive of schlieren setups, it does capture the governing physics. In this case, a point-light source is assumed, while in general, an extended light source would be used which would require a focusing lens to be positioned between the knife-edge and the screen \citep{Settles2001}.

The utility of a schlieren image lies in the fact that the refractive index is dependent on the density of the medium it passes through. Knowledge of this relationship permits one to obtain quantitative data from such images. To make this lucid, let $\rho \left( x, y \right)$ be the spatially varying density of a fluid surrounding an object of interest; it is expressed as a scalar-field with a horizontal coordinate $x$ and a vertical coordinate $y$. It has a linear relationship with the refractive index $n \left(x, y \right)$, given by

\begin{ceqn}
	\begin{align}
		n\left( x, y \right) - 1 = \kappa \rho \left( x, y \right)
		\label{equ:gdale}
	\end{align}
\end{ceqn}
where $\kappa$ is the Gladstone-Dale coefficient. A schlieren image represents the refractive index gradient field $\nabla n$, where the gradient direction is based on the orientation of the knife edge: placement in the vertical direction yields $\partial n / \partial x$, whilst placement along the horizontal direction renders $\partial n / \partial y$ \citep{hargather2012comparison}. 

There are two points to note regarding the overall quality of a schlieren image. First, the diffraction of light from the object of interest can degrade the quality of the schlieren image. This may be exacerbated around the edges of the test object where halos are found to appear. However, these do not introduce a significant impediment when used for quantitative analysis as they do not obscure the flow detail \citep{Kumar2008}. Second, is the prominence of dark regions in the schlieren image where it should be bright---this is suggestive of a region with high density gradients, where the light rays are highly refracted and blocked by some component in the schlieren set-up such as the knife-edge/lens mount \citep{Settles2001}. This may limit the accuracy of quantitative (and indeed qualitative) information that may be extracted from the region and its vicinity.

As mentioned previously, schlieren experiments have been conducted for a wide range of flows since its inception, however, it should be noted that it can be difficult to find an image pair corresponding to a vertical and horizontal knife-edge result in literature, with the vast majority of published results focusing on only one of the two knife-edge orientations. Additionally, the quality of published images can vary due to a wide range of reasons. One commonly seen issue with such images is over-ranging of the schlieren system, where too much knife-edge cutoff is used which leads to large areas showing up as extreme bright and dark regions, concealing the underlying flow features \citep{Settles2001}. Another commonly found issue is with older publications, where papers have been digitised, but with the pixel values of source images being subject to thresholding - resulting in much of the flow details being lost. While there have been a large number of schlieren images generated in literature, it is important to note their quality can vary widely due to the aforementioned reasons. 

Consider some of the earlier work conducted at the National Aeronautics and Space Administration (NASA) (or more specifically its predecessor the National Advisory Committee for Aeronautics (NACA)) where a significant number of schlieren results have been digitised, but without the same quality as a digitally captured image due to the digitisation process. With either significant noise being introduced, or the image pixel values being curtailed resulting in missing flow features and details. Some examples of such images can be found in the following references \citep{Daley1948,Lindsey1952,Erdmann1953,Hill1965}. It is important to note that such images would generate poor results with the proposed method.

\subsection{Supersonic flow behaviour} \label{sec:supersonic_flow}

\begin{figure}[!ht]
	\centering
	\includegraphics[width=0.5\columnwidth]{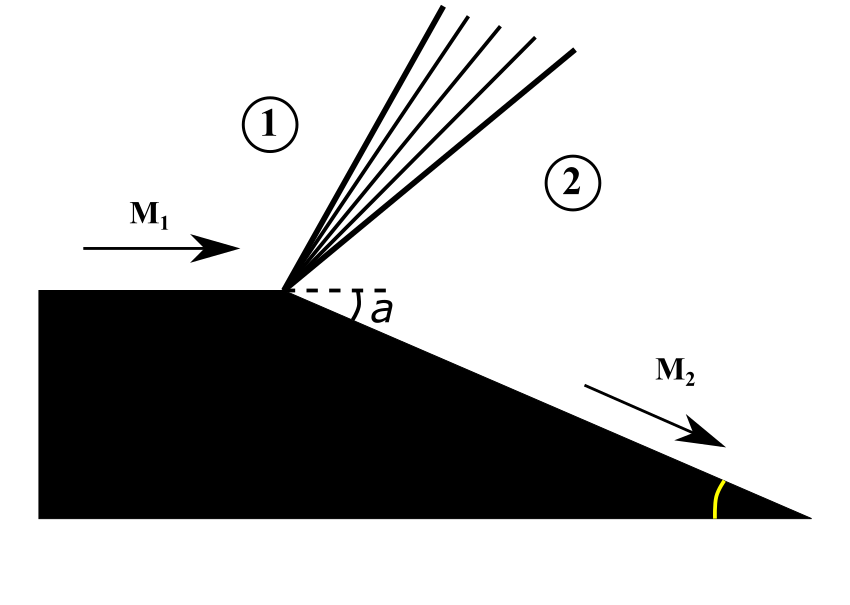}
	\caption{Expansion fan based on \cite{Anderson2011Fundamentals}}
	\label{fig:exp_fan}
\end{figure}

When an object is immersed in transonic or supersonic flow, thin regions of discontinuity can form across which flow properties change rapidly. Across these \emph{shock waves}, static flow properties such as pressure, density, temperature and entropy increase, while the stagnation pressure, Mach number and velocity decrease \citep{Anderson2011Fundamentals}. Oblique shock waves form across concave corners, where the shocks generally form at an oblique angle (less than $90^{\circ}$). In regions of convex corners, the opposite effect occurs, resulting in \emph{expansion fans}, across which the static pressure,  density, and temperature decrease, while the Mach number increases. This is shown in Figure \ref{fig:exp_fan}, where the flow follows the path from \circled{1} to \circled{2} across an expansion fan with an expansion angle $a$.

\begin{figure}[!ht]
	\begin{center}
		\begin{subfigmatrix}{2}
			\subfigure[]{\includegraphics[]{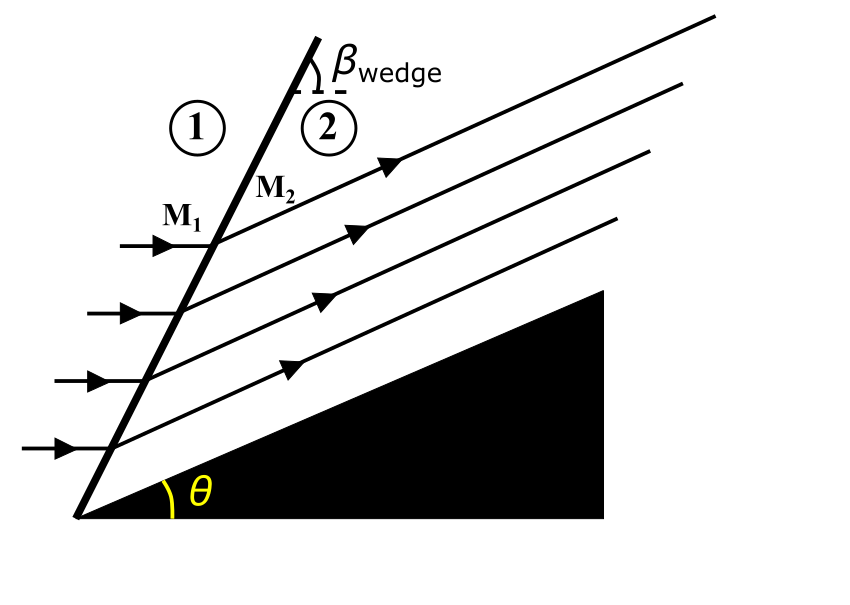}}
			\subfigure[]{\includegraphics[]{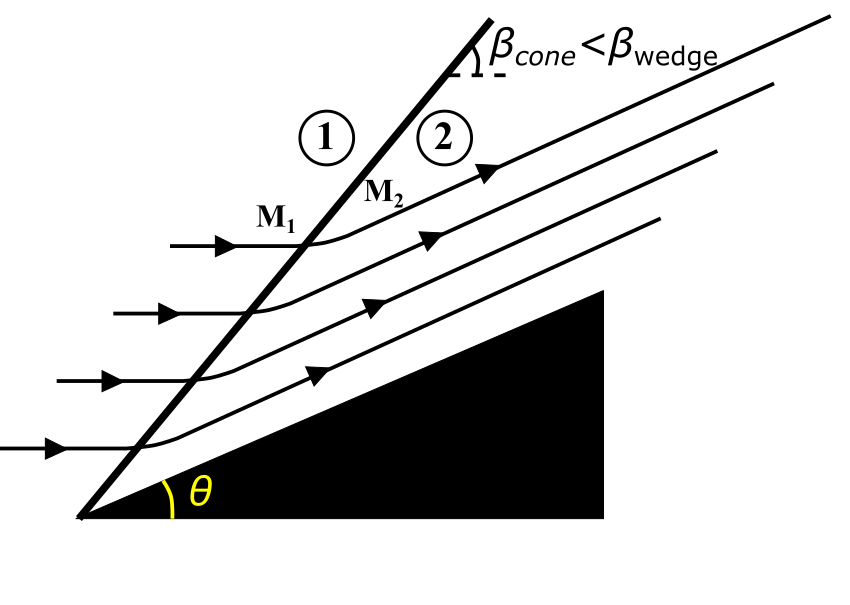}}
		\end{subfigmatrix}
		\caption{Oblique shocks in (a) wedge (2D), and (b) conical flow (3D), based on \cite{Anderson2011Fundamentals}}
		\label{fig:shocks}
	\end{center}
\end{figure}

As the flow must be tangential to the wall at all times, the streamlines will follow the direction of the wall. When considering a two-dimensional supersonic flow over a wedge, a sharp change in the direction of the streamline occurs across the shock. However, when considering three-dimensional supersonic flow over a cone, there is a third dimension for the air to pass through, leading to a ``three-dimensional relieving effect'' that renders a smoother transition to the new wall angle (see Figure \ref{fig:shocks}). This has the added effect of creating a weaker shock wave than a wedge shock with the same wall angle ($\theta$) and inflow properties. Hence, $\beta_{cone}<\beta_{wedge}$ under the same conditions. Apart from the weaker shock and the less adverse streamlines, the flow behaviour is similar between the wedge and the cone.

Assuming a steady state shock in a predominantly inviscid flow (except near the shock) under adiabatic conditions, the downstream Mach number and density across an oblique shock wave for a wedge, is given by (see page 623 in \cite{Anderson2011Fundamentals})

\begin{ceqn}
	\begin{equation}
		\begin{split}
			M_2 &= \frac{1}{sin(\beta_{wedge} - \theta)}\sqrt{\frac{1 + \frac{\gamma-1}{2}M_{1}^2 sin^2{\beta_{wedge}}}{\gamma M_{1}^2 sin^2{\beta_{wedge}} - \frac{\gamma-1}{2}}} \\ \\
			\rho_{2} &= \rho_{1} \frac{(\gamma + 1)M_{1}^2 sin^2{\beta_{wedge}}}{(\gamma - 1)M_{1}^2 sin^2{\beta_{wedge}} + 2}
		\end{split}
		\label{eqn:shock_relations}
	\end{equation}
\end{ceqn}

\noindent where, $\beta_{wedge}$ is the wave angle, $\theta$ is the deflection angle, $\gamma$ is the specific heat ratio, $M_{2}$ and $\rho_{2}$ are the downstream Mach and density. These expressions, in tandem with other isentropic flow relations, may be used to estimate density from other observed thermodynamic quantities.

\subsection{Quantitative analysis} \label{sec:quantitative_analysis}
Although predominantly used for qualitative assessments, there is a considerable body of literature devoted to the singular task of extracting quantitative information from schlieren images \citep{hargather2012comparison, wildeman2018real, hay2019sampling, tobin2016quantitative, venkatakrishnan2005density, dalziel2000whole}. While a typical qualitative analysis may study the visible shock angles and refraction patterns, a quantitative analysis focuses on extracting primitive flow field variables---e.g., density, temperature or velocity from schlieren images. The overarching framework relies on being able to relate the pixel values across the image to a known refraction angle \citep{Settles2001, Settles2017}. We remark that our definition of quantitative in this study refers strictly to the extraction of the density scalar field from schlieren images.

\subsubsection{Image pre-processing}
Depending on the schlieren set-up, images may require some level of pre-processing to account for poor contrast, brightness, and sharpness (to name a few factors) before use in quantitative analysis. The pixel-values of a schlieren image do not necessarily always encompass the entire 256 levels of greyscale possible; instead the brightest pixel may be a fraction of this. By setting the value of the brightest pixel to the highest greyscale level of 256, and proportionally increasing the value of the rest of the pixels, we can form a new image with a higher contrast. This is known as contrast stretching. The brightness of an image can be increased by simply increasing all the pixel values by a constant, with a maximum limit set to 256. The sharpness of the image can be improved using sharpening techniques such as unsharp masking. This method uses a mask created by the difference between the source image and a blurred version of the image to detect edges. The contrast of the edges covered by the mask are increased to create a sharper image (See Section 3.2.4 in \cite{Settles2001} for pre-processing techniques for schlieren images). 

\subsubsection{Calibrated Schlieren}
The most straightforward approach to extract quantitative data from a schlieren image is to place a calibrated object with a known refractive index in the test area. This would facilitate the generation of a map between the pixel intensity and the refractive index gradient. Typically, a lens is used, with each radial position across the lens having a different refraction angle (or refractive gradient) to focus the collimated light. The focal length of the calibration lens determines the resolution of the refractive index variations, with finer resolutions being obtainable with longer lenses. Imaging using this setup yields the calibration required between the pixel intensity and the true refractive index gradient. The refractive index gradient field is integrated to obtain the refractive index which is converted to a density field using the Gladstone-Dale relationship from \eqref{equ:gdale}, with a known density value used to solve for the constant of integration (typically the known free-stream density). This assumes the fluid under consideration is not a mixture of gases, and hence has a constant value of $\kappa$. To be able to quantify for a mixture of gases, an estimate for the Gladstone-Dale constant, $\kappa$ would be required \citep{Settles2001, Settles2017}.

\subsubsection{Rainbow Schlieren} 
Rainbow schlieren (discovered by \cite{Schardin1942, Vandiver1974}, but termed by Howes \citep{Howes1983, Howes1984}) substitutes a knife-edge with a radial rainbow cut-off filter with a continuous spectrum to generate an image with hue variations. The light passing through the colour cut-off filter will take the hue corresponding to its path along the filter. As the rainbow filter is continuous---i.e., without any sharp edges or discontinuities---errors due to diffraction are effectively non-existent. The replacement of the filter leads to refractive index gradients being displayed as variations in colour rather than irradiance \citep{hargather2012comparison}. Use of the colour filter by itself does not provide vastly more information over the standard greyscale schlieren image. However, its advantage lies in the ability to use different colour filters to colour-code targeted refraction directions/magnitudes. For example, the use of a bulls-eye pattern would highlight the refraction magnitudes that occur in the radial direction. Obtaining quantitative data from this setup would follow the described calibration approach from above, but instead of the pixel intensity of a greyscale schlieren image, the hue from a colour schlieren image would be mapped to the known refraction angle \citep{Settles2001, Settles2017}.

\subsubsection{Background Oriented Schlieren (BOS)}
The premise of BOS is to measure the level of distortion in a background pattern with and without the refractive disturbances. This method requires little set-up and equipment as only a high resolution camera is required, and possibly a background pattern depending on whether the natural background behind the object of interest has sufficient variation for image processing to pick up patterns and image distortion artefacts \citep{hargather2012comparison, Heineck2016, Heineck2021}. \cite{venkatakrishnan2005density} demonstrated the construction of a density field using BOS for a cone-cylinder flow at Mach 2.0. Additionally, work conducted by NASA has led to the use of natural backgrounds to capture schlieren images of different aircraft by using the sun/sky as the background \citep{Heineck2019, Hill2017a, Hill2017b}.

To estimate the density field using BOS, a reference image of a structured background pattern without density gradients is taken, along with another image with the refractive disturbances active. The displacements along $x$ and $y$ in the background pattern are computed between these two images using cross-correlation, and used to estimate a source term $S\left(x, y \right)$, which is the sum of the derivative of the density gradients. The individual density derivatives can then be obtained by substituting the empirical values of $S$ in Poisson's equation
\begin{ceqn}
	\begin{equation}
		\frac{\partial^2}{\partial x^2} \rho(x,y) + \frac{\partial^2}{\partial y^2} \rho(x,y) = S\left(x, y \right)
	\end{equation}
	\label{eqn:poisson}
\end{ceqn}
with appropriate boundary conditions (see 6 in \cite{venkatakrishnan2004density}). Note that this provides a projection of the 3D density field in the camera direction, \textit{i.e.} it is the integrated density distribution across the line of sight. By using standard tomographic reconstruction techniques such as filtered back-projection, the density field across a plane of interest can be obtained \citep{venkatakrishnan2005density}.

More quantitative methods such as absolute and standard photometry methods are described in detail in \cite{Settles2001}, Chapter 10. We omit an exposition of these methods here for brevity. It is important to note that all prior quantitative methods are limited, i.e., they require either a special apparatus with a reference calibration within the image, or multiple images of the same object with control over a refractive distortion. This makes it particularly difficult to re-create density fields of past schlieren images---especially those that were taken with only a qualitative analysis in mind. Additionally, none of these methods offer a comprehensive pathway to account for the uncertainties in the density (or other thermodynamic quantities).

\section{Pixels to Density via Machine Learning} \label{sec:methods}
Our approach to schlieren imaging in this paper takes a decidedly different path from the quantitative efforts referenced above. We leverage ideas from machine learning, namely Gaussian process regression \citep{rasmussen2006}, to facilitate the spatial estimation of density.

\subsection{From schlieren to density} \label{sec:sch_to_den}
Our goal in this subsection is to derive a formulation that maps one, or two, schlieren images---of the same observation taken with two different knife-edge orientations---to a spatial density estimate. A natural starting point is \eqref{equ:gdale}, that relates the refractive index with density. Taking partial derivatives with respect to $x$ and $y$ yields
\begin{align}
    \frac{\partial n\left(x, y \right) }{\partial x} & = \kappa \frac{\partial \rho \left(x, y \right)}{\partial x}, \; \; \; \text{and} \label{equ:grad_gdale_1}\\
    \frac{\partial n\left(x, y \right) }{\partial y} & = \kappa \frac{\partial \rho \left(x, y \right)}{\partial y}.
    \label{equ:grad_gdale_2}
\end{align}
The left-hand side of \eqref{equ:grad_gdale_1} and \eqref{equ:grad_gdale_2} is the partial derivative of the refractive index. Pixel intensities $\in \left[0, 255\right]$ in a given schlieren image are linearly related to their corresponding refractive index partial derivatives. In other words, a schlieren image taken with a vertical knife-edge is a function of $\partial \rho / \partial x$, and a schlieren image taken with a horizontal knife-edge is a function of $\partial \rho / \partial y$. The simplest way to characterise this linear relationship is
\begin{align}
    s_{x} & = \left( \frac{\frac{\partial n}{\partial x} - \textrm{min}\left( \frac{\partial n}{\partial x} \right) }{\textrm{max}\left( \frac{\partial n}{\partial x} \right) - \textrm{min}\left( \frac{\partial n}{\partial x} \right)} \right)c_{1} + c_{2}, \label{equ:schlieren_grad_1} \; \; \; \text{and} \\
    s_{y} & = \left( \frac{\frac{\partial n}{\partial y} - \textrm{min}\left( \frac{\partial n}{\partial y} \right) }{\textrm{max}\left( \frac{\partial n}{\partial y} \right) - \textrm{min}\left( \frac{\partial n}{\partial y} \right)} \right)c_{3} + c_{4}, \label{equ:schlieren_grad_2} 
\end{align}
where $s_x$ and $s_y$ are the pixel intensities in a schlieren image. Note that all the terms on the right-hand side of \eqref{equ:schlieren_grad_1} and \eqref{equ:schlieren_grad_2} barring $\partial n / \partial y$ and $\partial n / \partial x$ are constants. This includes $\left\{c_1, c_2, c_3, c_4 \right\}$, which will vary depending on the brightness, contrast, and other \emph{histogram-based} image processing choices. To clarify, there may be cases where both $c_1$ and $c_3$ will be 255, leading to both $c_2$ and $c_4$ to be zero.

Replacing partial derivatives of the refractive indices with the right-hand side of \eqref{equ:grad_gdale_1} and \eqref{equ:grad_gdale_2}, we obtain
\begin{align}
   s_{x} & = \left(\frac{\kappa\frac{\partial \rho}{\partial x} - \textrm{min}\left( \frac{\partial n}{\partial x} \right) }{\textrm{max}\left( \frac{\partial n}{\partial x} \right) - \textrm{min}\left( \frac{\partial n}{\partial x} \right)} \right)c_{1} + c_{2}, \label{equ:schlieren_grad_3} \; \; \; \text{and} \\
    s_{y} & = \left( \frac{\kappa \frac{\partial \rho}{\partial y} - \textrm{min}\left( \frac{\partial n}{\partial y} \right) }{\textrm{max}\left( \frac{\partial n}{\partial y} \right) - \textrm{min}\left( \frac{\partial n}{\partial y} \right)} \right)c_{3} + c_{4}, \label{equ:schlieren_grad_4}. 
\end{align}
As the constants above are not known a priori, we utilise a simpler characterisation
\begin{align}
    s_x= \alpha_x \frac{\partial \rho}{\partial x}  + \beta_x, \; \; \; \textrm{and} \; \; \; 
	s_y= \alpha_y \frac{\partial \rho}{\partial y}  + \beta_y,
	\label{equ:schlieren_density_linear}
\end{align}
where the model parameters $\left\{\alpha_x, \alpha_y, \beta_x, \beta_y\right\}$ need to inferred.

\subsection{Schlieren-based Gaussian process model}
We now build a Gaussian process \citep{rasmussen2006} model to statistically infer the density given one or two schlieren images. To ease our exposition, we enumerate the model \emph{training} and the model \emph{prediction} data below, where the abbreviation ``loc.'' denotes locations and ``out.'' denotes outputs.

\begin{table*}[h]
    \centering
    \caption{Data for the Schlieren-based Gaussian process model.}
    \begin{adjustbox}{max width=\textwidth}
    \begin{tabular}{p{20mm}lcccc}
    \textbf{Quantity} & \textbf{Training loc.} & \textbf{Training out.} & \textbf{Training noise} & \textbf{Prediction loc.} & \textbf{Prediction out.} \\[8pt]
    Schlieren \newline (vertical) & $\left\{ x^{\left(i\right)}_{\circ}, y^{\left(i\right)}_{\circ} \right\}_{i=1}^{N}$ & $\left\{ s_{x,\circ}^{\left(i\right)} \right\}_{i=1}^{N}$ & $\boldsymbol{\Sigma}_{x} \in \mathbb{R}^{N\times N}$ & $\left\{ x^{\left(i\right)}_{\ast}, y^{\left(i\right)}_{\ast} \right\}_{i=1}^{M}$ & $\left\{ s^{\left(i\right)}_{x,\ast} \right\}_{i=1}^{M}$\\[8pt]
    Schlieren  \newline (horizontal) & $\left\{ x^{\left(i\right)}_{\circ}, y^{\left(i\right)}_{\circ} \right\}_{i=1}^{N}$ & $\left\{ s_{y,\circ}^{\left(i\right)} \right\}_{i=1}^{N}$ & $\boldsymbol{\Sigma}_{y} \in \mathbb{R}^{N\times N}$ & $\left\{ x^{\left(i\right)}_{\ast}, y^{\left(i\right)}_{\ast} \right\}_{i=1}^{M}$ & $\left\{ s^{\left(i\right)}_{y,\ast} \right\}_{i=1}^{M}$\\[8pt]
    Density & $\left\{ x^{\left(i\right)}_{\bullet}, y^{\left(i\right)}_{\bullet} \right\}_{i=1}^{P}$ & $\left\{ \rho_{\bullet}^{\left(i\right)} \right\}_{i=1}^{P}$ & $\boldsymbol{\Psi} \in \mathbb{R}^{P\times P}$ & $\left\{ x^{\left(i\right)}_{\dagger}, y^{\left(i\right)}_{\dagger} \right\}_{i=1}^{D}$ & $\left\{ \rho^{\left(i\right)}_{\dagger} \right\}_{i=1}^{D}$\\[8pt]
    \end{tabular}
    \end{adjustbox}
    \label{tab:notation}
\end{table*}

\begin{figure}[!ht]
    \centering
    \includegraphics[width=0.95\textwidth]{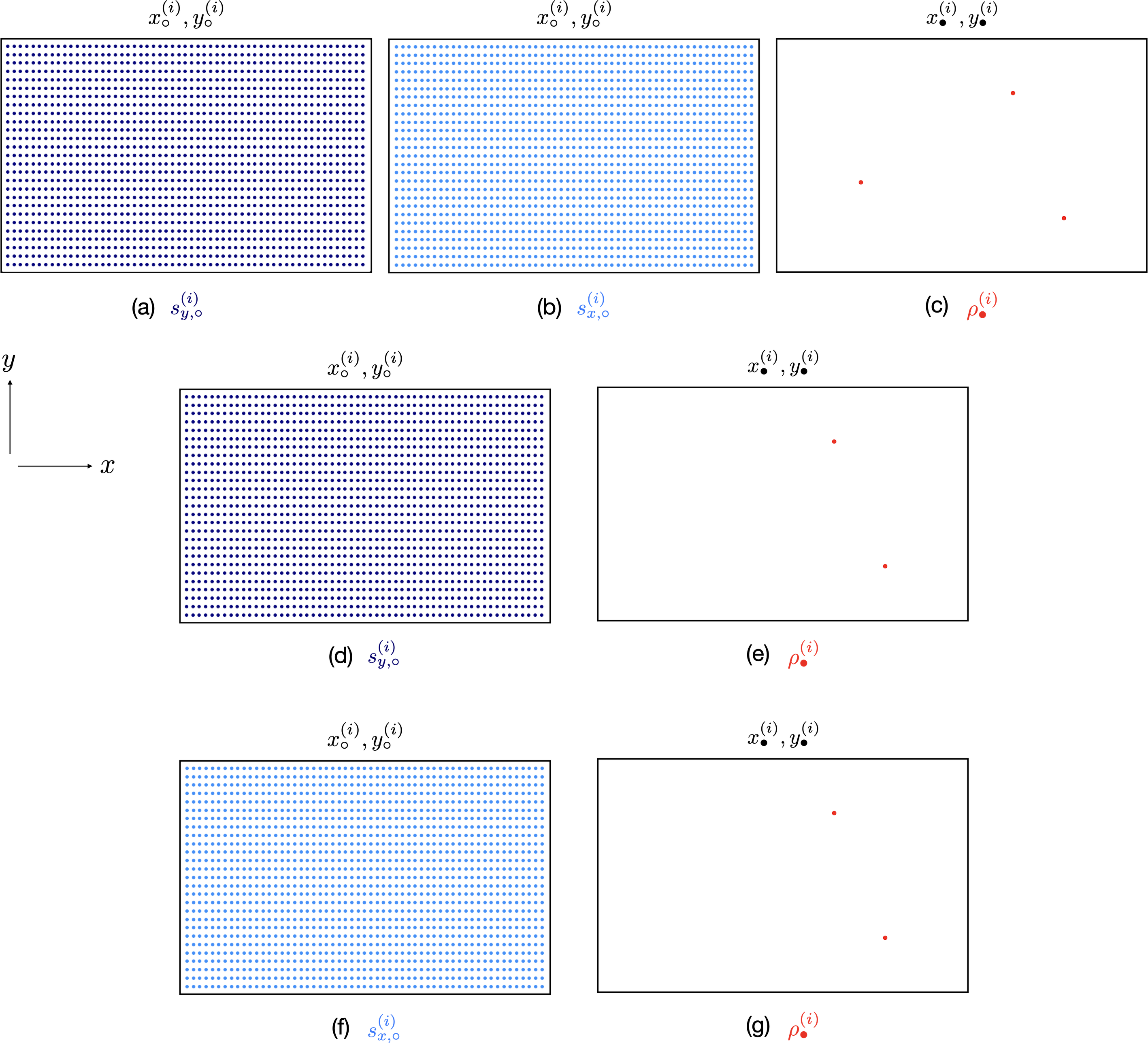}
    \caption{Possible model inputs spatial location configurations. Rows (a-c) represent the case with inputs from two schlieren images and three density values; rows (d-e) and (f-g) represent the case with one schlieren image and two density values}
    \label{fig:model_inputs}
\end{figure}

To crystallise the precise form of the model inputs, consider the schematic shown in Figure~\ref{fig:model_inputs}. Each row represents a possible combination of model inputs. In the top row, the grid of points represent training locations taken from both (a) horizontal (b) and vertical schlieren images, along with (c) three density spatial locations. Sub-figures (d-e) represent spatial samples associated with only the (d) horizontal schlieren image and (e) two density input locations. Lastly sub-figures (f-g) show the (f) vertical schlieren image and (g) two density input locations. As we will see later, the distinction between a model input with one or two schlieren images will have effect the output spatial density estimates. We remark here that while the input spatial locations for the schlieren images in Figure~\ref{fig:model_inputs} are shown as a uniform grid, in practice one can resort to non-uniform sampling with the aim of delivering a better resolution of certain spatial characteristics.

In table~\ref{tab:notation}, each row has an associated \emph{training noise} that attributes an uncertainty to the pixel and density values. This uncertainty is assumed to have a zero-mean Gaussian distribution. Throughout this paper, we further assume that these covariance matrices have the form $\boldsymbol{\Sigma}_{x} = \sigma_{s_x}^2 \mathbf{I}$, $\boldsymbol{\Sigma}_{y} = \sigma_{s_y}^2 \mathbf{I}$, and $\boldsymbol{\Psi} = \sigma_{\rho}^2 \mathbf{I}$, where $\mathbf{I}$ is the identity matrix and standard deviations $\sigma_{s_x}, \sigma_{s_y}$ and $\sigma_{\rho}$ will be assigned.

It is important to note that the quality of the schlieren images does have a direct impact on the reliability of the result generated using this method. Source images where there are regions with thresholded values, clipping or large amounts of noise in general (\textit{e.g.} due to digitisation) are not good candidates for this method since it can produce non-physical or poor results.

\subsubsection{Model prior}
Standard Bayesian formalism requires the definition of a model likelihood and a model prior. The model likelihood is set by the data and the noise, while the model prior in a Gaussian process context is set by the mean and covariance function. To this end, we assume that the spatial density field---which we seek to infer---bounded by the image domain is a Gaussian random field with some mean and covariance function. Were we to select a few spatial samples $\left(x, y\right)$ from this field, they would represent a Gaussian distribution specified entirely by the mean and covariance \citep{rasmussen2006}. 

Note from \eqref{equ:schlieren_density_linear}, schlieren images are described as linear operators acting over the density field. By assuming that the density field is a Gaussian random field, and that the schlieren images are effectively linear operators acting over this random field, a gradient-based Gaussian process framework for extracting the density can be derived. The \emph{joint} Gaussian density across the training data (see first three columns of Table~\ref{tab:notation}) is written as
\begin{equation}
\underbrace{\left[\begin{array}{c}
\left(s_{x,\circ}^{\left(1\right)},\ldots,s_{x,\circ}^{\left(N\right)}\right)^{T}\\
\left(s_{y,\circ}^{\left(1\right)},\ldots,s_{y,\circ}^{\left(N\right)}\right)^{T}\\
\left(\rho_{\dagger}^{\left(1\right)},\ldots,\rho_{\dagger}^{\left(P\right)}\right)
\end{array}\right]}_{\mathbf{f}} \sim\mathcal{N}\left(\left[\begin{array}{c}
-\beta_{x} \cdot \boldsymbol{e}_{N} \\
-\beta_{y} \cdot\boldsymbol{e}_{N}\\
0 \cdot \boldsymbol{e}_{P}
\end{array}\right], \underbrace{\left[\begin{array}{ccc}
\alpha_{x}^{2}\nabla_{x^{2}}\mathbf{K}_{\circ\circ}+\boldsymbol{\Sigma}_{x} & \alpha_{x}\alpha_{y}\nabla_{xy}\mathbf{K}_{\circ\circ} & \alpha_{x}\nabla_{x}\mathbf{K}_{\circ\bullet}\\
\alpha_{x}\alpha_{y}\nabla_{yx}\mathbf{K}_{\circ\circ} & \alpha_{y}^{2}\nabla_{y^{2}}\mathbf{K}_{\circ\circ}+\boldsymbol{\Sigma}_{y} & \alpha_{y}\nabla_{y}\mathbf{K}_{\circ\bullet}\\
\left(\alpha_{x}\nabla_{x}\mathbf{K}_{\circ\bullet}\right)^{T} & \left(\alpha_{y}\nabla_{y}\mathbf{K}_{\circ\bullet}\right)^{T} & \mathbf{K}_{\bullet\bullet}+\boldsymbol{\Psi}
\end{array}\right]}_{\mathbf{C}}\right),
\label{equ:joint_dist_gp_modified}
\end{equation}
where $\boldsymbol{e}_{\left( \cdot \right) }$ denotes a vector of ones of length specified by its subscript. Matrices given by the symbol $\mathbf{K}$ are used to denote a \emph{kernel function} $k\left(\left(x, y\right), \left(x', y'\right) \right)$ evaluated at the spatial locations provided as the arguments. Element-wise entries of the relevant blocks of the covariance matrix $\mathbf{C}$ in \eqref{equ:joint_dist_gp_modified} are given by
\begin{align}
\begin{split}
    \left[\mathbf{K}_{\bullet\bullet}\right]_{ij} = k\left(\left(x^{\left(i\right)}_{\bullet}, y^{\left(i\right)}_{\bullet}\right), \left(x^{\left(j\right)}_{\bullet}, y^{\left(j\right)}_{\bullet}\right)\right), & \; \; \; \left[\nabla_{x}\mathbf{K}_{\circ\bullet}\right]_{ij} = \frac{\partial k\left(\left(x^{\left(i\right)}_{\circ}, y^{\left(i\right)}_{\circ}\right), \left(x^{\left(j\right)}_{\bullet}, y^{\left(j\right)}_{\bullet}\right)\right)}{\partial x}, \\
    \left[\nabla_{y}\mathbf{K}_{\circ\bullet}\right]_{ij} = \frac{\partial k\left(\left(x^{\left(i\right)}_{\circ}, y^{\left(i\right)}_{\circ}\right), \left(x^{\left(j\right)}_{\bullet}, y^{\left(j\right)}_{\bullet}\right)\right)}{\partial y}, & \; \; \; \left[\nabla_{x}^2\mathbf{K}_{\circ\bullet}\right]_{ij} = \frac{\partial^2 k\left(\left(x^{\left(i\right)}_{\circ}, y^{\left(i\right)}_{\circ}\right), \left(x^{\left(j\right)}_{\bullet}, y^{\left(j\right)}_{\bullet}\right)\right)}{\partial x^2}, \\
    \left[\nabla_{y}^2\mathbf{K}_{\circ\bullet}\right]_{ij} = \frac{\partial^2 k\left(\left(x^{\left(i\right)}_{\circ}, y^{\left(i\right)}_{\circ}\right), \left(x^{\left(j\right)}_{\bullet}, y^{\left(j\right)}_{\bullet}\right)\right)}{\partial y^2} & \; \; \; \left[\nabla_{xy}\mathbf{K}_{\circ\bullet}\right]_{ij} = \frac{\partial^2 k\left(\left(x^{\left(i\right)}_{\circ}, y^{\left(i\right)}_{\circ}\right), \left(x^{\left(j\right)}_{\bullet}, y^{\left(j\right)}_{\bullet}\right)\right)}{\partial x \partial y}, \\
    \left[\nabla_{yx}\mathbf{K}_{\circ\bullet}\right]_{ij} = \frac{\partial^2 k\left(\left(x^{\left(i\right)}_{\circ}, y^{\left(i\right)}_{\circ}\right), \left(x^{\left(j\right)}_{\bullet}, y^{\left(j\right)}_{\bullet}\right)\right)}{\partial y \partial x}, & \; \; \; \textrm{and} \; \; \; \left[\mathbf{K}_{\circ\circ}\right]_{ij} = k\left(\left(x^{\left(i\right)}_{\circ}, y^{\left(i\right)}_{\circ}\right), \left(x^{\left(j\right)}_{\circ}, y^{\left(j\right)}_{\circ}\right)\right).
    \label{equ:covariance_blocks}
\end{split}
\end{align}
It is important to note that $\mathbf{C}$ in \eqref{equ:joint_dist_gp_modified} must be symmetric positive definite for the chosen kernel function $k$. The predictive mean for the two schlieren images and the density may then be computed as
\begin{equation} \label{equ:gek_mean}
\mathbb{E}\left[\begin{array}{c}
\left(s_{x,\ast}^{\left(1\right)},\ldots,s_{x,\ast}^{\left(N\right)}\right)^{T}\\
\left(s_{y,\ast}^{\left(1\right)},\ldots,s_{y,\ast}^{\left(N\right)}\right)^{T}\\
\left(\rho_{\dagger}^{\left(1\right)},\ldots,\rho_{\dagger}^{\left(P\right)}\right)
\end{array}\right]=\left[\begin{array}{c}
\nabla_{x}\mathbf{K}_{\ast\circ}\\
\nabla_{y}\mathbf{K}_{\ast\circ}\\
\mathbf{K}_{\dagger\bullet}
\end{array}\right]\mathbf{C}^{-1}\mathbf{f}
\end{equation}
with $\mathbf{f}$ representing the stacked training data outputs as per \eqref{equ:joint_dist_gp_modified}. The predictive covariance is written as
\begin{equation} \label{equ:gek_covar}
\mathbb{C}\mathrm{ovar}\left[\begin{array}{c}
\left(s_{x,\ast}^{\left(1\right)},\ldots,s_{x,\ast}^{\left(N\right)}\right)^{T}\\
\left(s_{y,\ast}^{\left(1\right)},\ldots,s_{y,\ast}^{\left(N\right)}\right)^{T}\\
\left(\rho_{\dagger}^{\left(1\right)},\ldots,\rho_{\dagger}^{\left(P\right)}\right)
\end{array}\right]=\left[\begin{array}{c}
\nabla_{x}\mathbf{K}_{\ast\ast}\\
\nabla_{y}\mathbf{K}_{\ast\ast}\\
\mathbf{K}_{\dagger\dagger}
\end{array}\right]-\left[\begin{array}{c}
\nabla_{x}\mathbf{K}_{\ast\circ}\\
\nabla_{y}\mathbf{K}_{\ast\circ}\\
\mathbf{K}_{\dagger\bullet}
\end{array}\right]\mathbf{C}^{-1}\left[\begin{array}{c}
\nabla_{x}\mathbf{K}_{\ast\circ}\\
\nabla_{y}\mathbf{K}_{\ast\circ}\\
\mathbf{K}_{\dagger\bullet}
\end{array}\right]^{T}
\end{equation}
where the individual covariance matrix blocks follow the same notation in \eqref{equ:covariance_blocks} trading the \emph{training location} subscripts with the \emph{prediction locations} as per Table~\ref{tab:notation}.

A few comments on the above are in order. First, our model is distinct from the standard gradient enhanced Gaussian process model \citep{sarkka2011linear} (also known as Gradient-Enhanced Kriging \citep{Lockwood2012,Ulaganathan2015,Ulaganathan2016,Wu2017,Chen2019,Bouhlel2019}) as we are not strictly observing gradients, but some linear function of the gradients following \eqref{equ:schlieren_density_linear}. Second, while there is considerable flexibility in defining the kernel function, it must be twice differentiable. One obvious choice is the squared exponential function 
\begin{equation}
    k\left(\left(x, y\right), \left(x', y'\right) \right) = \sigma_{f}^2 \;  \textrm{exp}\left( -\frac{1}{2} \left[ \frac{\left(x- x' \right)^2 }{l_{x}^2} + \frac{\left(y- y' \right)^2 }{l_{y}^2} \right] \right) 
\end{equation}
that is parameterised by a signal noise variance $\sigma_{f}$ and a length scales $l_x$ and $l_y$. Expressions for 
\begin{equation}
    \frac{\partial k}{\partial x}, \; \; \; \frac{\partial k}{\partial y}, \; \; \; \frac{\partial^2 k}{\partial x^2}, \; \; \; \frac{\partial^2 k}{\partial y^2}, \; \; \; \frac{\partial k}{\partial x \partial y}, \; \; \; \text{and} \; \; \; \frac{\partial k}{\partial y \partial x}
\end{equation}
are trivially derived and used in \eqref{equ:covariance_blocks}. Other valid kernel choices include a Fourier series kernel \citep{seshadri2022}, a Mat\'{e}rn kernel \citep{rasmussen2006}, among others.

A subset of the joint distribution defined in equation \eqref{equ:joint_dist_gp_modified} can be used if an image is available for only one gradient direction with the caveat that the posterior mean will not fully represent the density field unless the flow is highly directional and the gradient in the other direction is negligible.

\subsubsection{Prior definition}
Unknown parameters associated with both the kernel function and the overarching model can be grouped into a vector  $\boldsymbol{\theta}=\left(\alpha_x, \alpha_y, \beta_x, \beta_y, l_x, l_y, \sigma_f\right)$. By incorporating knowledge of the potential range or physical nature of the model, it is possible to prescribe a set of prior probability distributions $p(\boldsymbol{\theta})$. If knowledge regarding the observed data is unknown, this can be captured by a prior distribution with a large variance. In this study, prior distributions are defined for $\boldsymbol{\theta}$ with reduced model complexity to allow for generalisation of the model to multiple cases. To ensure a positive semi-definite covariance matrix, the hyperparameters defined here must be positive. Additionally, to ensure that the length scales and kernel noise terms are not wildly different at the gradient- and functional levels, we introduce a re-parameterisation of the form
\begin{ceqn}
	\begin{align}
		\alpha_x = \frac{\lambda_x \lambda_y}{l_x^2},
		\nonumber \\
		\alpha_y = \frac{\lambda_x \lambda_y}{l_y^2},
		\nonumber \\
		\sigma_f^2 = \frac{\xi}{\lambda_x \lambda_y}.
	\end{align}
\end{ceqn}
\noindent where, $\lambda_x$, $\lambda_y$, $l_x$, and $l_y$ are assigned half-normal prior distributions. With this new prior definition, we set $\boldsymbol{\theta} = \left( \xi, \lambda_x, \lambda_y, l_x, l_y, \beta_x, \beta_y \right)$. When observation data for only schlieren image is available, the hyperparameters simplify to the original $\boldsymbol{\theta} = \left( \xi, \alpha_x, \alpha_y, l_x, l_y, \beta_x, \beta_y \right)$.

\subsubsection{Hyperparameter optimisation}
These free hyperparameters ($\boldsymbol{\theta}$) require tuning to find the optimal values to give the expected posterior mean. This can be via a straightforward grid or random search but with the downside of being intractable when working with a larger number of hyperparameters \citep{Claesen2015}. There are instead automated methods that approach this search from a more analytical perspective.

A well-worn approach is to maximise the logarithm of the marginal likelihood, conditioned on the hyperparameters \citep{rasmussen2006}. The values found from such a method are such that they maximise the likelihood that the user-specified model produces the data that is observed.

\begin{ceqn}
	\begin{eqnarray}
		\underset{\boldsymbol{\theta}}{\textrm{maximize}} \; \; \; \textrm{log} \left( \textrm{p} \right) := && -\frac{1}{2} \boldsymbol{\mu}_{\tilde{\rho}}^{T}\mC_{\tilde{\rho}}^{-1}\boldsymbol{\mu}_{\tilde{\rho}} - \frac{1}{2} \textrm{log}\left|\mC_{\tilde{\rho}} \right| 
		- \frac{N}{2} \textrm{log}\left( 2 \pi \right),
		\label{equ:maxlike}
	\end{eqnarray}
\end{ceqn}

where the mean and covariance terms are evaluated at the observed data only. A more developed approach is the Maximum a Posteriori (MAP) estimation method which in addition to the log-likelihood ($\textrm{log} \; p\left(x \mid \boldsymbol{\theta} \right)$) takes advantage of knowledge in the prior distribution with the $\textrm{log} \; p(\boldsymbol{\theta})$ term. The prior distribution creates a bias of the probability mass density towards regions that are preferred a priori \citep{Goodfellow2016}, which can be advantageous when trusted prior knowledge of the observed data is available.

\begin{ceqn}
	\begin{eqnarray}
		\boldsymbol{\theta}_{\textrm{MAP}} \;\; = \underset{\boldsymbol{\theta}}{\textrm{arg\ max}} \; p\left( \boldsymbol{\theta} \mid x \right) = && \underset{\boldsymbol{\theta}}{\textrm{arg\ max}} \; \textrm{log} \; p\left( x \mid \boldsymbol{\theta} \right) 
		+ \textrm{log} \; p \left( \boldsymbol{\theta} \right).
		\label{equ:map}
	\end{eqnarray}
\end{ceqn}

Both the marginal likelihood and MAP make predictions based on a point estimate of $\theta$ which can be advantageous from a performance standpoint, but have the side-effect of focusing on a local maxima and not widely exploring the probability space. Ideally, a fully Bayesian approach would be used where instead of a point estimate, predictions would be made with a full distribution over $\boldsymbol{\theta}$, where, for each observed sample, either a positive probability would contribute for the next sample and any uncertainty would be accounted for in any predictions made \citep{Goodfellow2016}. However, it is intractable to attempt to make predictions using a full posterior distribution over $\boldsymbol{\theta}$.

One approach to achieve an estimated result is by approximating the posterior distribution by random sampling. A popular method is Markov chain Monte Carlo (MCMC). MCMC can also be used to generate an estimate of the posterior distribution where it cannot be calculated analytically. MCMC performs approximate integration by drawing samples from the posterior distribution and performing sample averages. The Markov chain refers to the drawing of samples, where, the next sample drawn is dependent on the previous, and Monte Carlo is the technique used for random sampling of the posterior distribution for integration \citep{Goodfellow2016}. By running multiple chains concurrently with varying starting points, it is possible to capture multiple states of equilibrium unlike the point estimate approaches.

\subsubsection{Computational Considerations}
Computing the direct inverse of covariance matrices in equations \ref{equ:gek_mean}, and \ref{equ:gek_covar} can be slow and numerically unstable. If one were to use a standard approaches, such as an LU decomposition to find the inverse, the cost would be in the order $\mathcal{O}(2n^3/3)$ for an n-dimensional matrix \citep{Gibbs1997}. By considering the symmetric and positive-definite nature of the covariance matrix, the Cholesky decomposition can instead be used which exploits the symmetry to cost half that of the LU \citep{Trefethen1997} factorisation. This cost can be intractable for larger datasets, especially when applying Bayesian approach to optimise for kernel hyperparameters ($\theta$), where $K$ is evaluated repeatedly with varying $\theta$. There are many approaches in literature that aim to solve this issue, either by creating low-rank approximations of $K$ at the cost of accuracy, by exploiting the sampling structure to break down the inversion to more efficient matrix vector products, or through iterative approximations such as the conjugate gradient method \citep{Gilboa2015, Shen2005, Gibbs1998, Skilling1993}.

One approach to reduce computational cost is through the use sparse GPs that approximate the posterior with pseudo-training points (known as inducing points, $m$) that are located within the domain. If the original number of training inputs is $n$, with $m$ < $n$, the original $\mathcal{O}(2n^3/3)$ cost of direct covariance matrix inversion is reduced down to $\mathcal{O}(nm^2)$. Two such commonly used sparse approximation methods are FITC (fully independent training conditional) \citep{Snelson2005, Candela2005}, and VFE (variational free energy) \citep{Titsias2009}. 

Work by \cite{Saatci2011} demonstrates the use of a grid structure for training points which results in a $Kronecker$ covariance matrix where Kronecker products can be used to efficiently complete Gaussian regression in $\mathcal{O}(n)$. Due to the non-collocated nature of the approach in this study (between the gradient and function sampling), the gradient inputs may be sampled from an image in a grid format, however the addition of the function observations disrupts the required format to take advantage of Kronecker products directly, without any masking of regions. Since this paper is purely focused on the methodology of obtaining quantitative data from schlieren images, only some of these approaches have been considered to make the method tractable.

\section{Results}
\begin{figure*}[ht!]
	\centering
	\includegraphics[width=\textwidth]{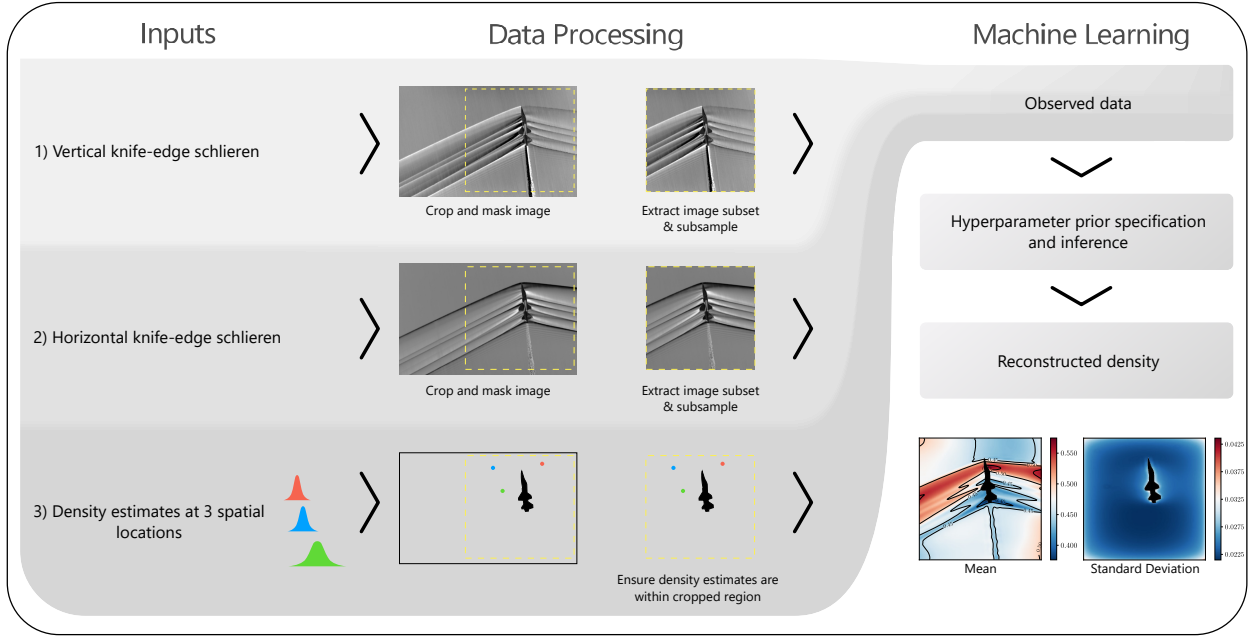}
	\caption{An overview of the machine learning framework for reconstructing density from schlieren images. Aircraft images shown are obtained from \cite{Heineck2021}}
	\label{fig:process_chart}
\end{figure*}

This paper offers a radically different way to think about schlieren, whilst staying true to its fundamental physics, building upon our preliminary work towards quantitative schlieren using machine learning \citep{ubald2021quantitative}. An overview of the present workflow is illustrated in Figure \ref{fig:process_chart}.
In the results that follow, we report the mean and standard deviation, i.e., $\text{sqrt}\left[ \textrm{diag}\left( \mathbb{C}\mathrm{ovar} \right) \right]$ output by our model. As these moments are for the density field, both have units of $kg/m^3$.

It should be noted that unless otherwise stated, that a down-sampled version of the image is used as input in each case due to the computational cost. Additionally, Mat\'{e}rn 3/2 and 5/2 kernel functions were also used to compare against the squared exponential kernel function defined in equation \ref{equ:covariance_blocks} for the first case, all kernel functions obtained comparable predictions of density. However, the Mat\'{e}rn kernels were more onerous to optimise for compared to the squared exponential, hence, for all studies in this paper, the squared exponential kernel function is used.

\subsection{Wind tunnel sting model}
\begin{figure}[ht!]
	\begin{center}
		\begin{subfigmatrix}{2}
			\subfigure[]{\includegraphics[]{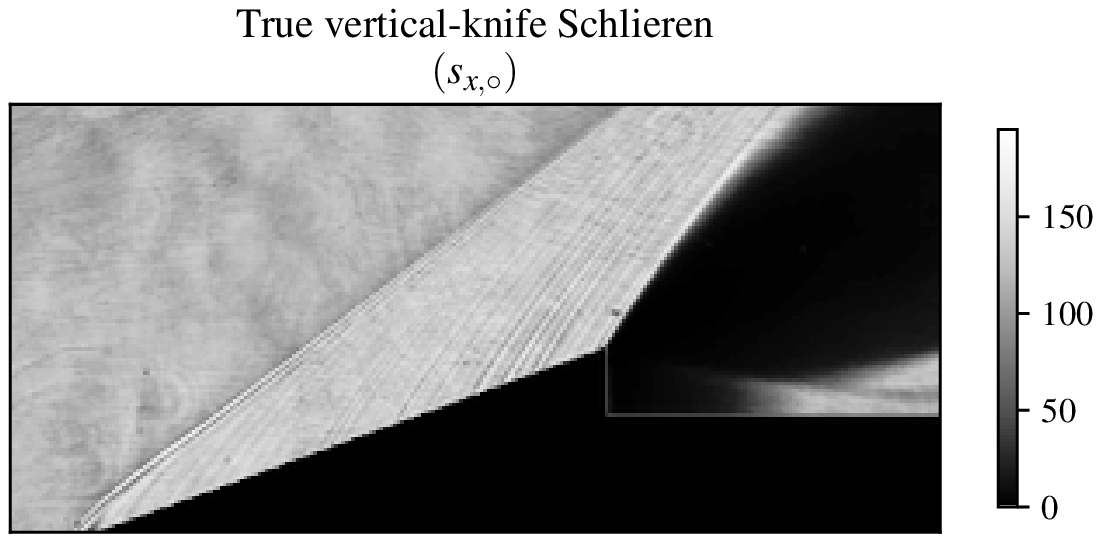}}
			\subfigure[]{\includegraphics[]{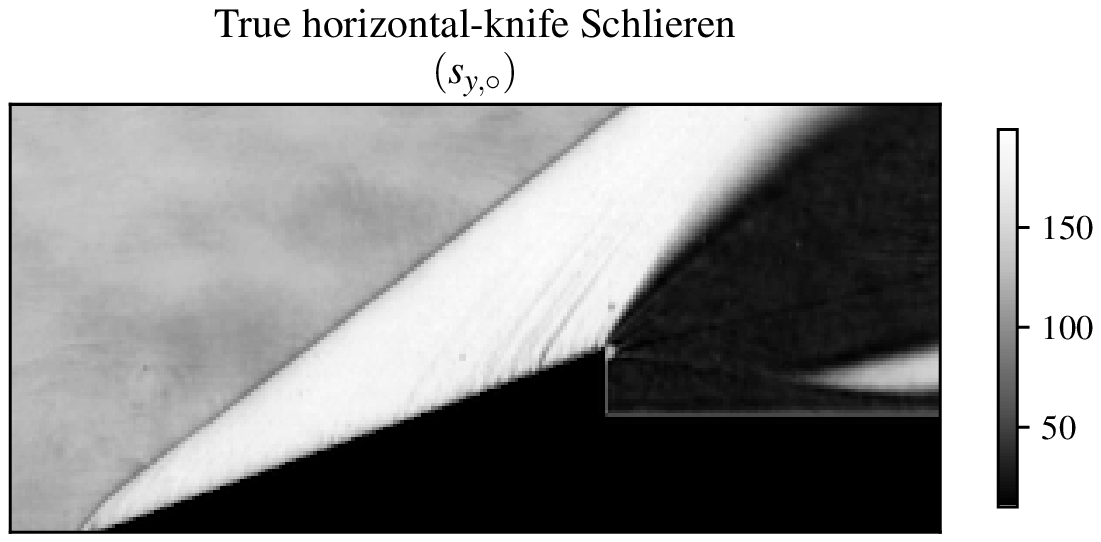}}
			\subfigure[]{\includegraphics[]{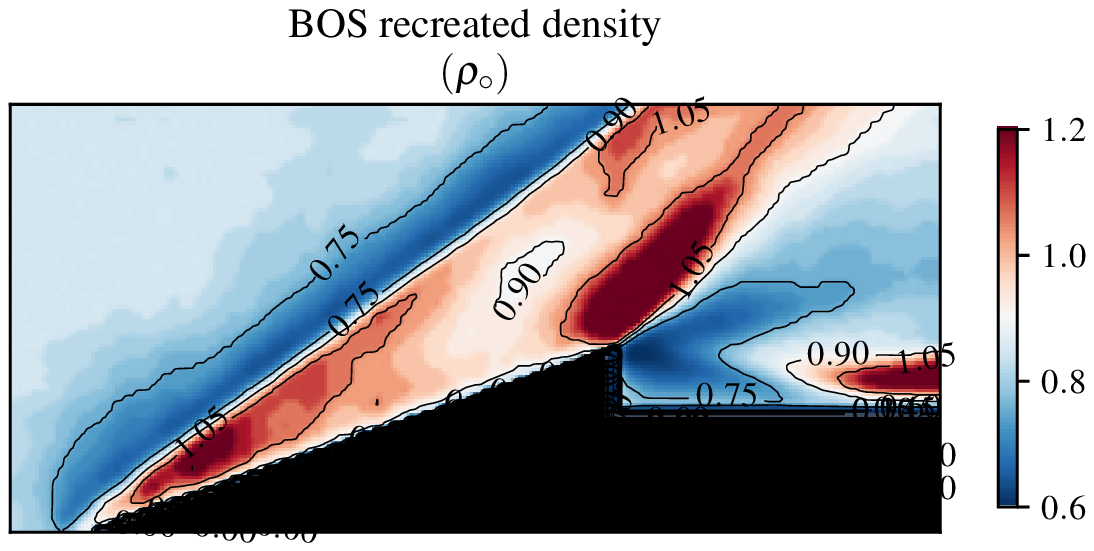}}
		\end{subfigmatrix}
		\caption{Experimental schlieren results from \cite{ota2011schlieren}: (a) vertical knife-edge schlieren image; (b) horizontal knife-edge schlieren image, and (c) BOS yielded density estimation. Data obtained from Masanori Ota.}
		\label{fig:sting_exp}
	\end{center}
\end{figure}

Consider the asymmetrical sting model in supersonic flow ($M_\infty =$ 2.0) published by \cite{ota2011schlieren}. The reconstructed density published in that paper, using BOS, is assumed to be representative of the \emph{true} density, and thus serves as our benchmark for comparison. Figure \ref{fig:sting_exp} captures the experimental results with the vertical knife-edge schlieren image in (a); horizontal knife-edge schlieren image in (b), and the BOS yielded density reconstruction in (c). Note that for subfigures (a) and (b) the colours denote greyscale pixel values between $[0,255]$, whilst the colours in (c) represent the density in units of $kg/m^3$. Another point to note is that only the upper half of the schlieren images are used for this reconstruction, as the shock captured across the lower half sees excessive clipping.

\begin{figure}[ht!]
	\begin{center}
		\begin{subfigmatrix}{2}
			\subfigure[]{\includegraphics[]{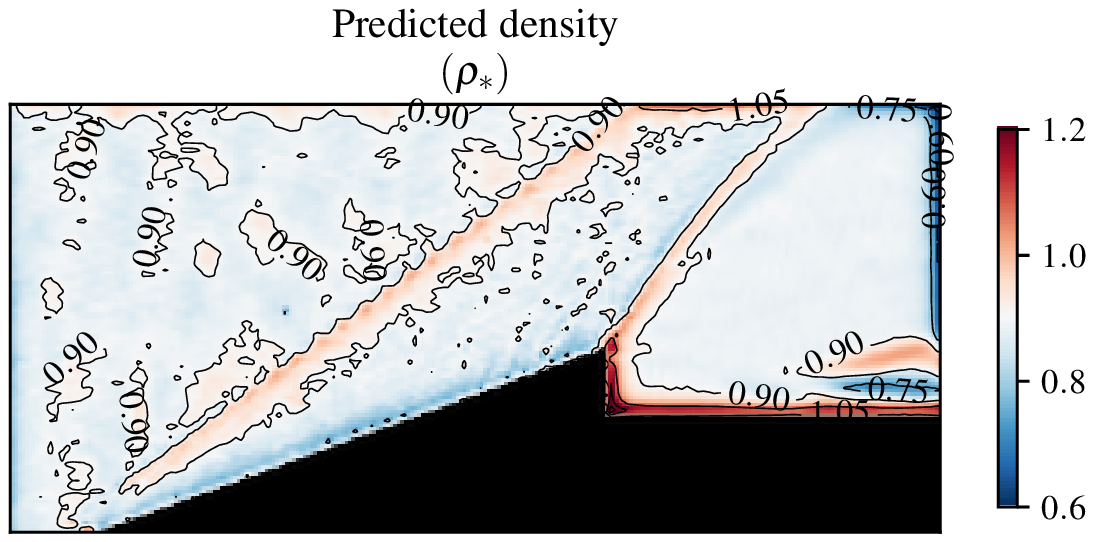}}
			\subfigure[]{\includegraphics[]{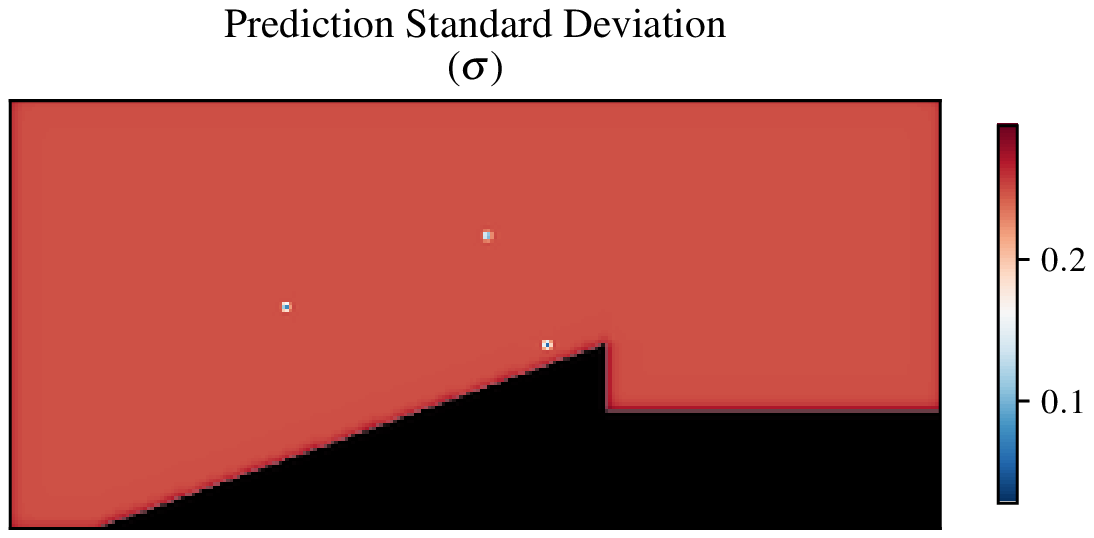}}
		\end{subfigmatrix}
		\caption{Density reconstruction from the vertical and horizontal knife-edge schlieren images of the sting, obtained as a Gaussian process with (a) mean; (b) standard deviation. Both are moments with units of $kg/m^3$}
		\label{fig:sting_gp_results}
	\end{center}
\end{figure}

The reconstructed density is shown in Figure \ref{fig:sting_gp_results} with the mean in (a) and the standard deviation in (b). The mean follows the expected trend of a uniform density upstream of the shock followed by a rapid increase in density across the shock, before decreasing across the expansion fan. Due to clipping of the input schlieren images, the gradients in region downstream of the expansion fan are not captured (the gradients are close to zero), hence this is represented as the mean of the observed densities instead. We remark that the BOS density in subplot Figure \ref{fig:sting_exp}(c) does not follow the expected behaviour of oblique shocks in theory, where, the density upstream of the shock must be uniform with a sharp increase in density across the shock. This may be caused by optical distortion incurred when performing BOS. Additionally, the shock itself should be captured as a very thin band, which is what is captured in our reconstruction, along with the shear layer downstream. 

\begin{figure}[ht!]
	\begin{center}
		\begin{subfigmatrix}{2}
			\subfigure[]{\includegraphics[]{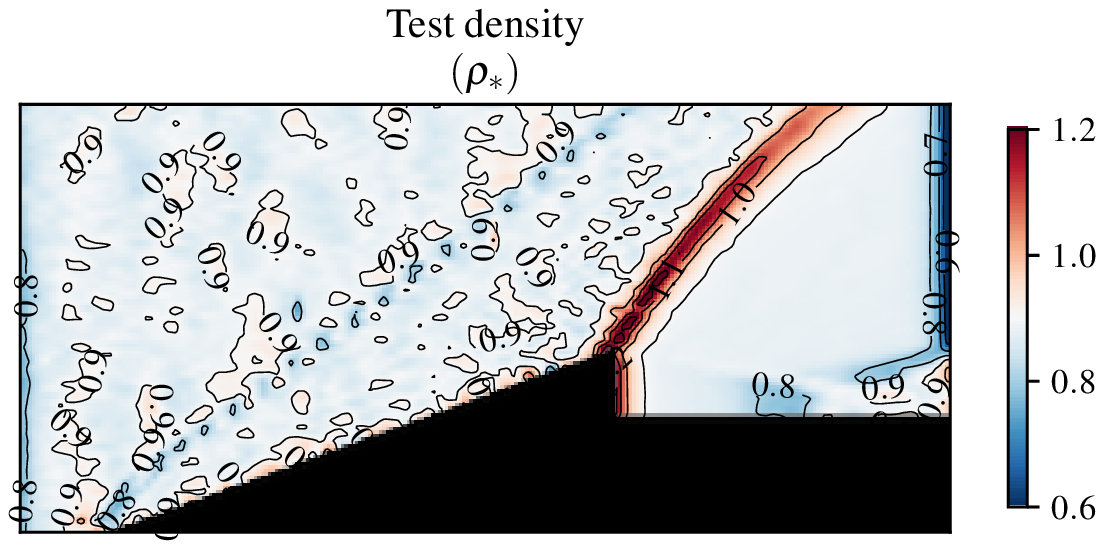}}
			\subfigure[]{\includegraphics[]{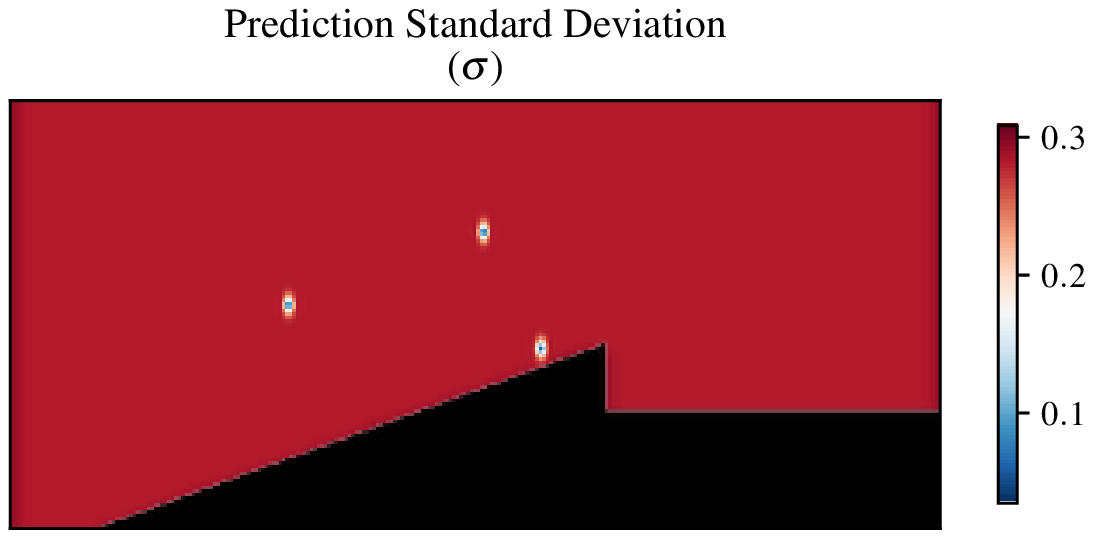}}
		\end{subfigmatrix}
		\caption{Density reconstruction from the vertical knife-edge schlieren image of the sting, obtained as a Gaussian process with (a) mean; (b) standard deviation. Both are moments with units of $kg/m^3$}
		\label{fig:sting_dx_gp_results}
	\end{center}
\end{figure}

It is possible to obtain some results using partial gradient observation, \textit{i.e.}, using either vertical or horizontal knife-edge schlieren as input. An example of the density prediction using the same workflow as above, but using only vertical knife-edge as input is shown in Figure \ref{fig:sting_dx_gp_results}(a). A semblance of the shock is formed with this approach, but is not fully formed since the density gradient is directional, but the shear layer downstream is captured. The lack of additional gradient input is captured by the higher standard deviation in the prediction (Figure \ref{fig:sting_dx_gp_results}(b)).

\FloatBarrier
\subsection{Supersonic aircraft in flight}
\begin{figure}[ht!]
	\begin{center}
	\begin{subfigmatrix}{2}
		\subfigure[]{\includegraphics[]{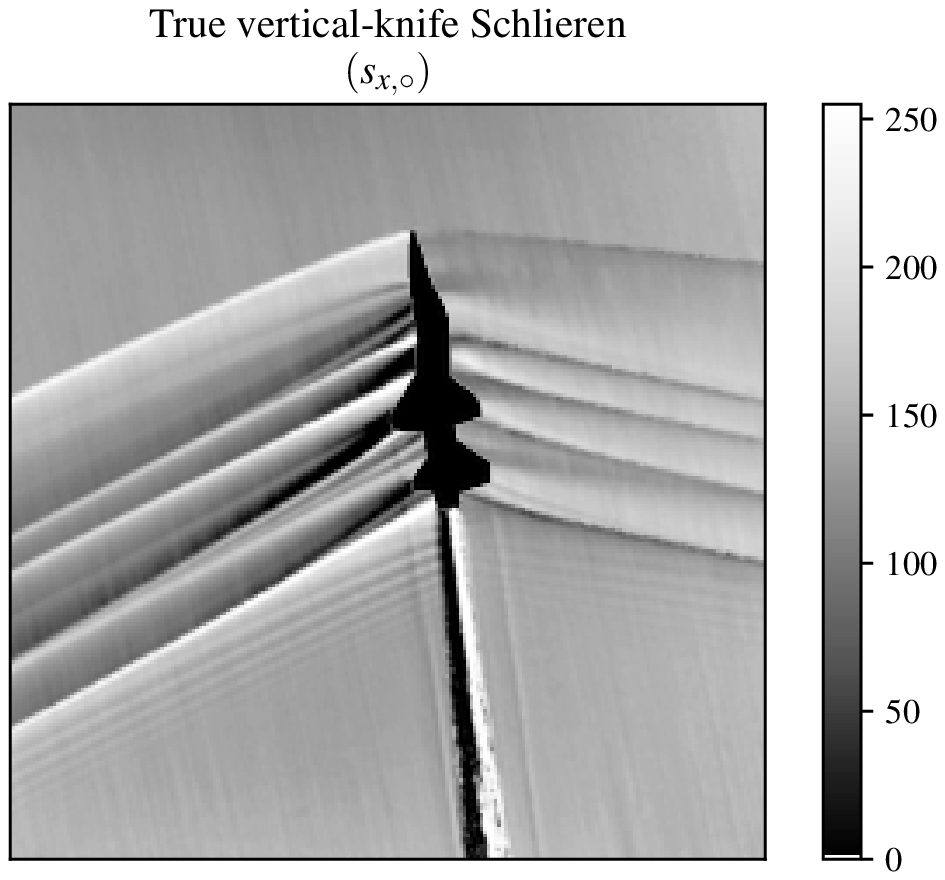}}
		\subfigure[]{\includegraphics[]{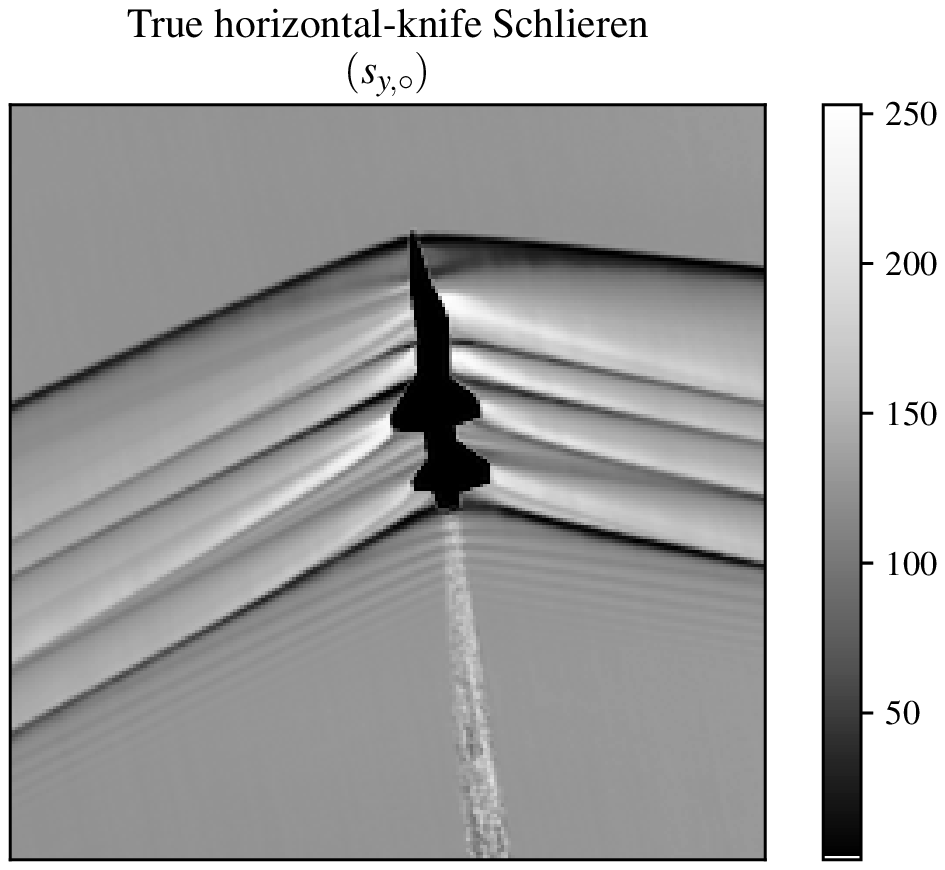}}
		\subfigure[]{\includegraphics[frame]{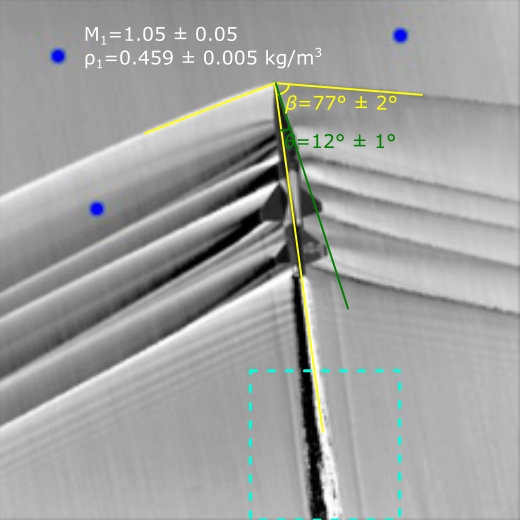}}
	\end{subfigmatrix}
	\caption{Experimental schlieren results from \cite{Heineck2016, Heineck2021}: (a) vertical knife-edge schlieren image, and (b) horizontal knife-edge schlieren image. Our estimate of some of the flow quantities in (c) and the uncertainties therein}
	\label{fig:nasa_aircraft}
	\end{center}
\end{figure}

Next, we study schlieren images taken by NASA of a T-38 training jet at a Mach number of 1.05. The horizontal and vertical knife-edge schlieren images captured by \cite{Heineck2016, Heineck2021} are shown in Figure \ref{fig:nasa_aircraft}(a) and (b) respectively. Additionally, based on estimates of the ambient conditions of the flight and estimates of shock and wedge angles, we can estimate the density at three points as shown in Figure~\ref{fig:nasa_aircraft}(c). The two freestream points are assigned the same value of density, whilst the density at the third (remaining) point is estimated via the shock relations in section \ref{sec:supersonic_flow}. 
 
As true density fields are not available for this case, and at least one known density is required from a non-freestream region for this method, the density after the first oblique shock wave is estimated. Two quantities are used to calculate the density: (i) the known experimental altitude of 30,000 ft and thus the freestream density; (ii) the shock relations in Equation \ref{eqn:shock_relations}, accounting for uncertainties in the Mach number and angles. However, note that this is a simplification, as the wedge angle calculation does not account for the three-dimensional relief effect. The flow conditions at this altitude and the estimated uncertainties are listed in the appendix in Table \ref{tab:case3_flowcons}. This density data, along with horizontal and vertical knife-edge schlieren images are sub-sampled (to reduce the computational overhead), and fed into the machine learning workflow. For the inputs to the GP, the aircraft region is masked out, as it is the geometry and is not part of the flow.

\begin{figure}[ht!]
	\begin{center}
		\begin{subfigmatrix}{2}
			\subfigure[]{\includegraphics[]{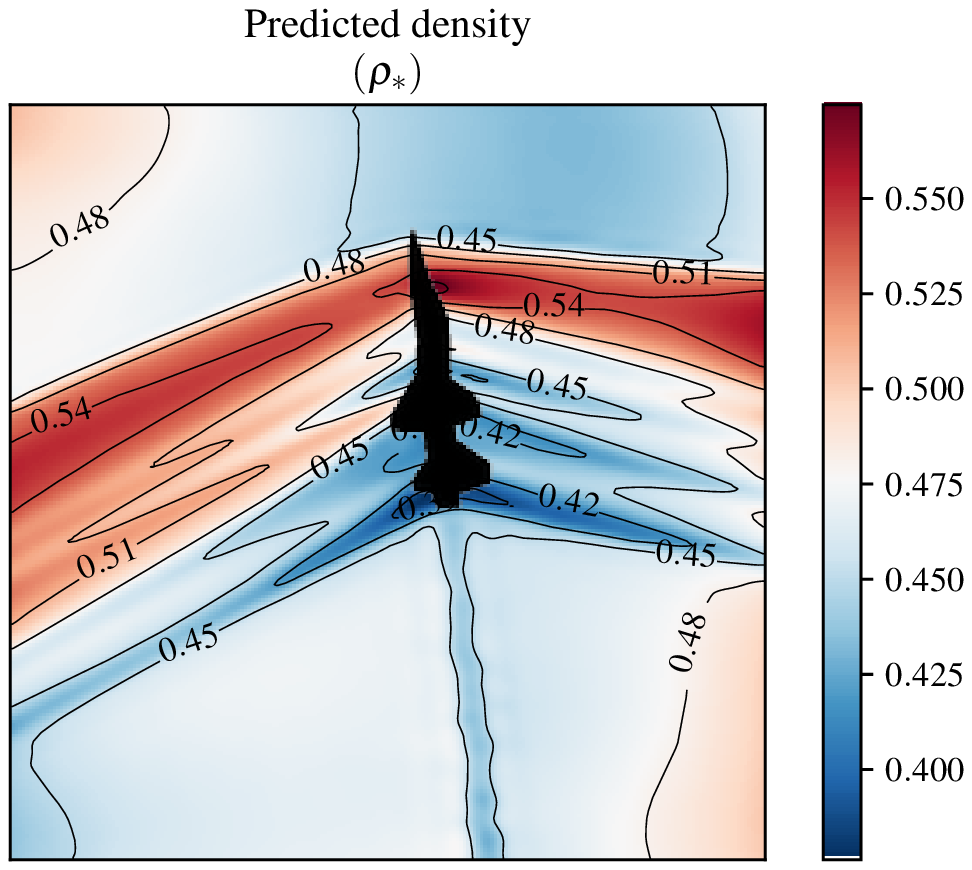}}
			\subfigure[]{\includegraphics[]{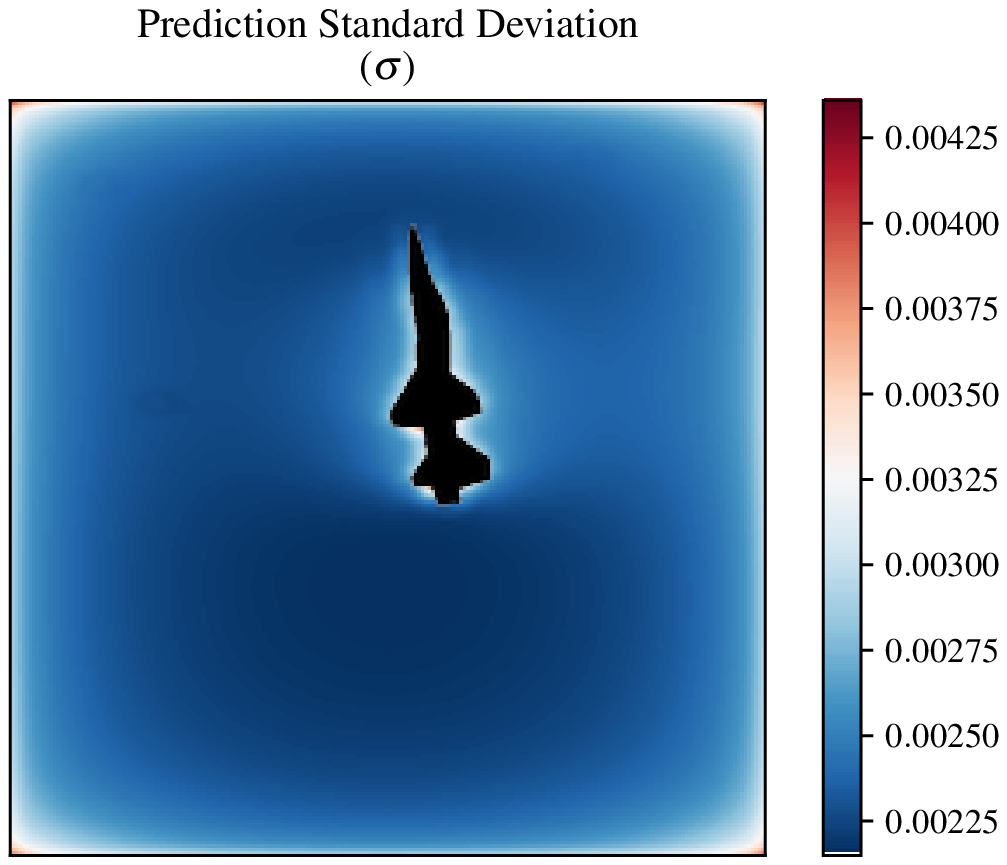}}
		\end{subfigmatrix}
		\caption{Density reconstruction from the vertical and horizontal knife-edge schlieren images of the supersonic aircraft, obtained as a Gaussian process with (a) mean; (b) standard deviation. Both are moments with units of $kg/m^3$}
		\label{fig:nasa_results}
	\end{center}
\end{figure}

The predicted density and standard deviation are shown in Figure \ref{fig:nasa_results}, with each shock clearly visible in the prediction. In addition to this, by focusing in on the downstream wake region (highlighted blue section in Figure \ref{fig:nasa_aircraft}(c)), the wake prediction can be seen in more detail, with the wake trail clearly being captured by the prediction.

\begin{figure}[ht!]
	\begin{center}
		\begin{subfigmatrix}{2}
			\subfigure[]{\includegraphics[]{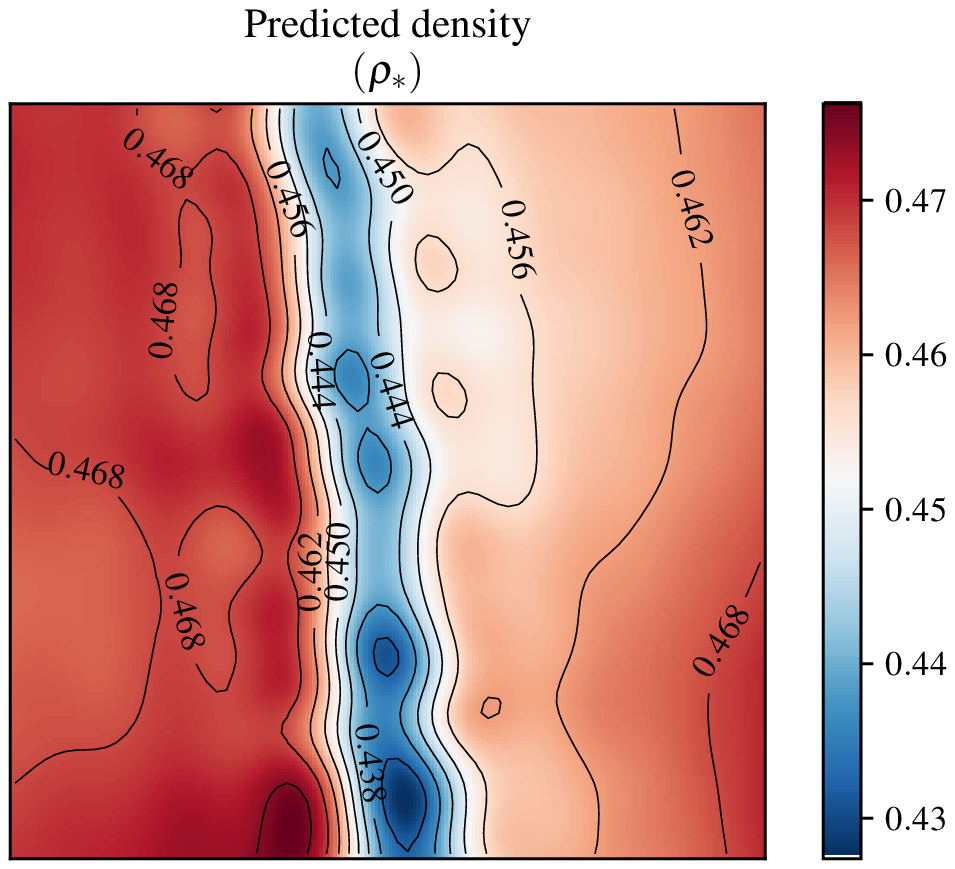}}
			\subfigure[]{\includegraphics[]{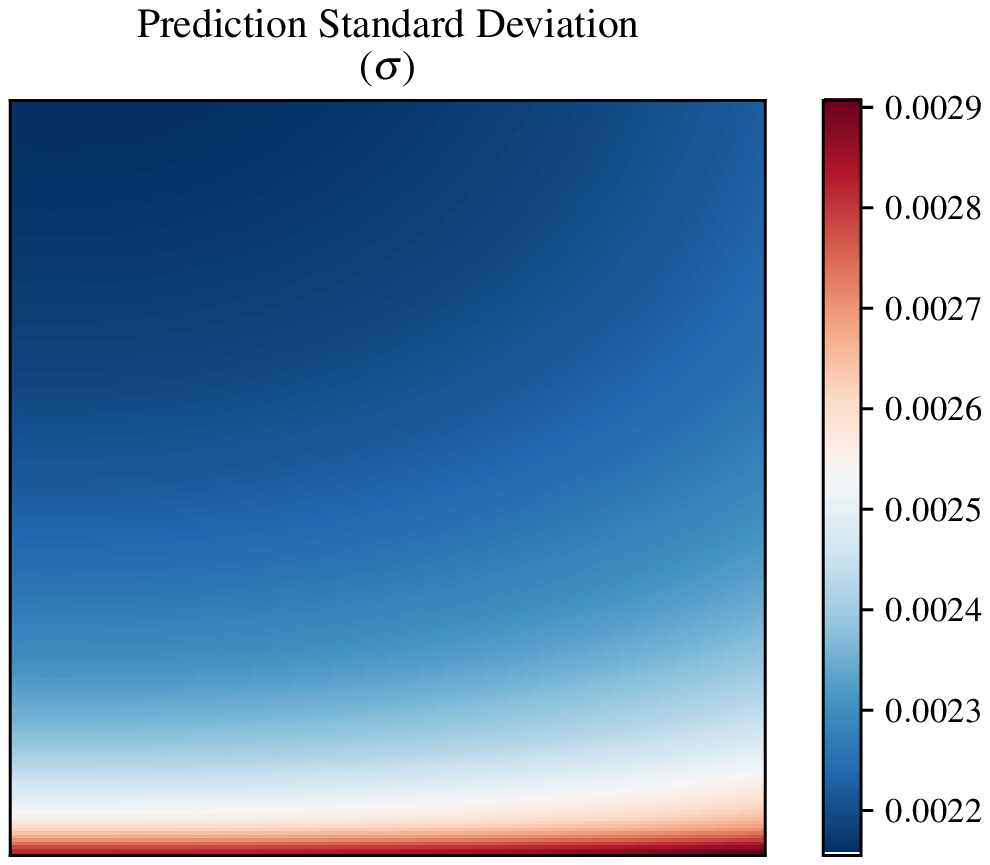}}
		\end{subfigmatrix}
		\caption{Density reconstruction from the vertical and horizontal knife-edge schlieren images of the supersonic aircraft, focusing on the downstream wake region, obtained as a Gaussian process with (a) mean; (b) standard deviation. Both are moments with units of $kg/m^3$}
		\label{fig:nasa_wake_results}
	\end{center}
\end{figure}

\FloatBarrier
\subsection{Shock tube CFD}
\begin{figure}[ht!]
	\begin{center}
		\begin{subfigmatrix}{2}
			\subfigure[]{\includegraphics[]{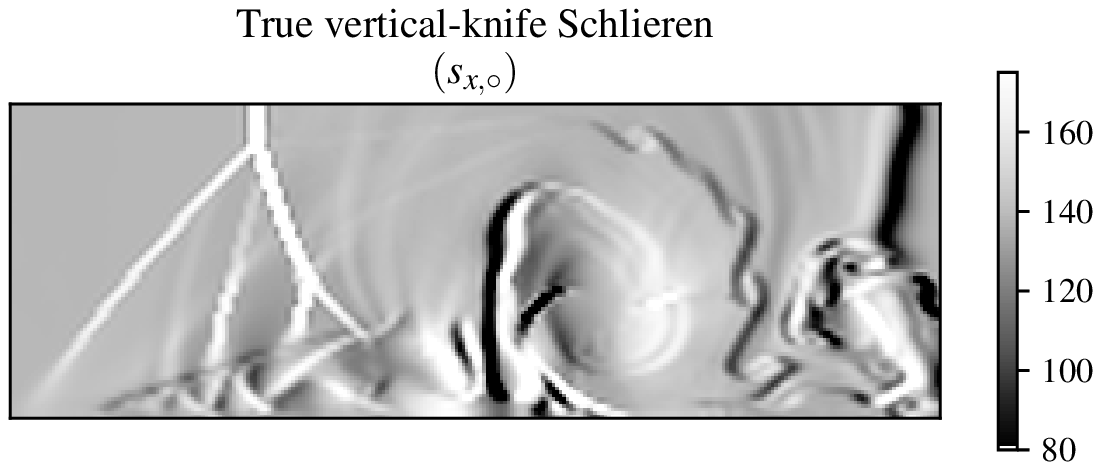}}
			\subfigure[]{\includegraphics[]{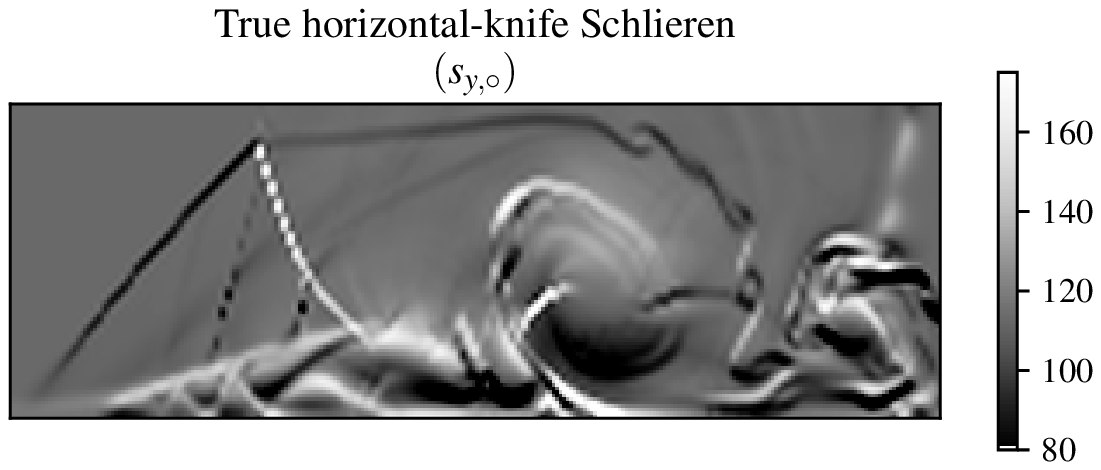}}
			\subfigure[]{\includegraphics[]{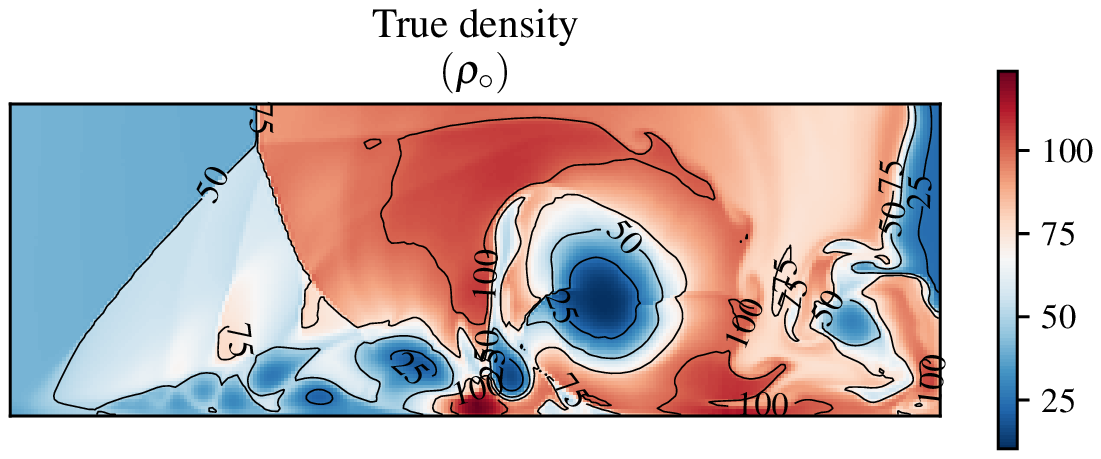}}
		\end{subfigmatrix}
		\caption{Numerical results of shock tube test case \citep{Dzanic2022}: (a) vertical knife-edge schlieren image; (b) horizontal knife-edge schlieren image, and (c) True density from CFD}
		\label{fig:shocktube_cfd}
	\end{center}
\end{figure}

In this section, we consider a canonical Sod shock tube \citep{Sod1978} test case. The CFD simulation result provided by \cite{Dzanic2022} is used, with the full primitive flow variables available for comparison. Figure \ref{fig:shocktube_cfd} shows a set of results from one timestep that are used as inputs to the GP. To simulate a schlieren image, the density gradient is calculated, and normalised to standard pixel value range of between 0 and 255. Figure \ref{fig:shocktube_cfd} shows the vertical knife-edge schlieren result in (a); horizontal knife-edge schlieren image in (b), and the true density obtained by CFD (c).

\begin{figure}[ht!]
	\begin{center}
		\begin{subfigmatrix}{2}
			\subfigure[]{\includegraphics[]{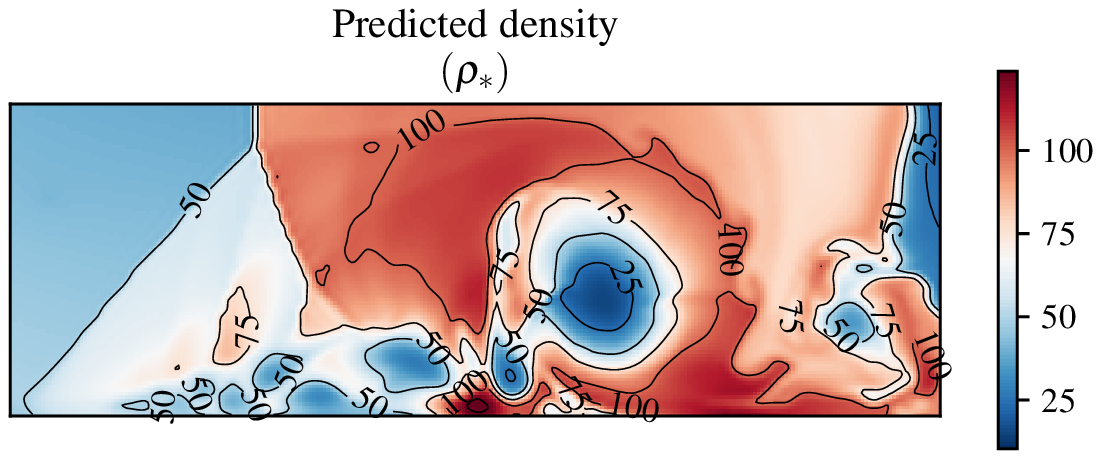}}
			\subfigure[]{\includegraphics[]{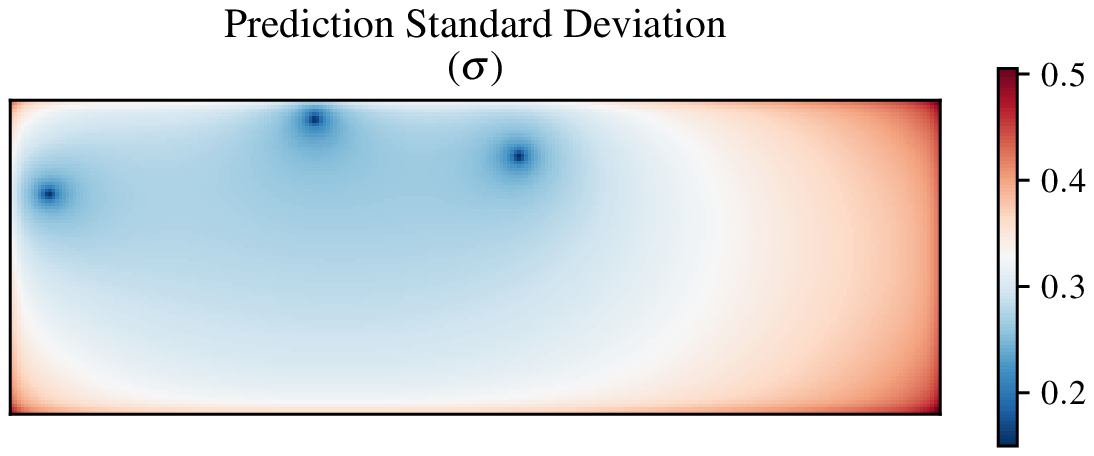}}
		\end{subfigmatrix}
		\caption{Density reconstruction from the vertical and horizontal knife-edge schlieren images of the shocktube case, obtained as a Gaussian process with (a) mean; (b) standard deviation. Both are moments with units of $kg/m^3$}
		\label{fig:shocktube_cfd_results}
	\end{center}
\end{figure}

\FloatBarrier
\section{Conclusions and Future Work}
We have demonstrated the development of a method to obtain quantitative density data from pre-existing Schlieren images through the use of Gaussian processes. The central novelty with this method lies in the fact that it does not require any pre-calibration or special set-up as required by existing methods discussed in Section~\ref{sec:quantitative_analysis}, and is widely applicable---even if only one knife-edge orientation is available. Our workflow is not intended to upend or subvert the modern practicalities of BOS, but rather complement it, by enabling scientists to be able to retrospectively reconstruct density fields from the plethora of schlieren images that exist in literature.

Future work will study the use of the above methodology for schlieren videos, which will incorporate a temporal kernel that relates successive frames of a video. Beyond this we plan to explore the use of deep Gaussian processes to facilitate more accurate flow feature representations.


\appendix
\section{Appendix}
In this section we present a validation of our approach along with supplemental particulars for the two examples provided above.

\subsection{Method validation with synthetic Schlieren images}
\begin{figure}[ht!]
	\centering
	\includegraphics[width=0.33\textwidth]{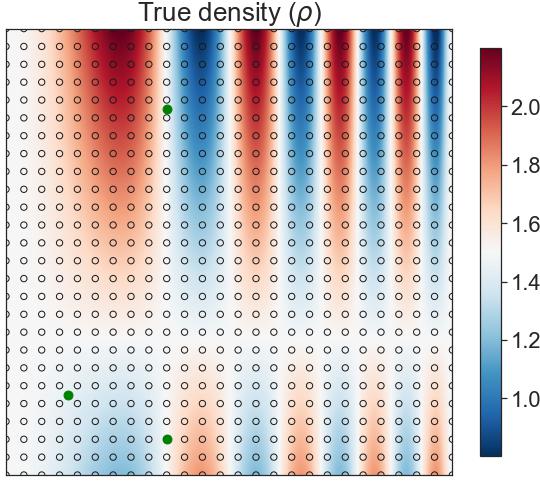}
	\caption{Analytical test case sampling locations}
	\label{fig:analytic_sampling}
\end{figure}

A simple analytical function is used to define a model density field and its corresponding density gradient. The density gradients are normalised to pixel values ($s^{\ast} \in \left[0, 255 \right]$) to emulate a Schlieren image with an unknown range, and a random sampling of 200 are selected from this image for use as observation data (white circles in Figure \ref{fig:analytic_sampling}), with three random density sampling locations also used (green circles in Figure \ref{fig:analytic_sampling}). The density and its partial derivative w.r.t $x$ and $y$ for this case are given as

\begin{ceqn}
	\begin{equation} \label{eqn:rank1}
		\begin{split}
			\rho &= sin( x^2 ) (2y - 3)/10 + 1.5, \\
			\frac{\partial{\rho}}{\partial{x}} &= x\ cos(x^2) (2y - 3)/5, \\
			\frac{\partial{\rho}}{\partial{y}} &= sin(x^2)/5.
		\end{split}
		\
	\end{equation}
\end{ceqn}

\begin{table}[ht!]
	\centering
	\caption{Prior definition for case 1}
	\label{tab:case1_params}
	\begin{tabular}{@{}llll@{}}
		\toprule
		Parameter & Distribution & Moments &  \\ \midrule
		$l_x, l_y$        &  half-normal  &     $\sigma=2.0$    &  \\
		$\lambda_x, \lambda_y$    & half-normal  &    $\sigma=1.0$     &  \\
		$\xi$      & half-normal  &    $\sigma=1.0$     &  \\
		$\sigma_y$    & deterministic  &    1E-3     &  \\
		$\sigma_{dy}$    & deterministic  &    0.1     &  \\ \bottomrule
	\end{tabular}
\end{table}

\begin{figure}[ht!]
	\begin{center}
		\begin{subfigmatrix}{3}
			\subfigure[]{\includegraphics[width=0.325\textwidth]{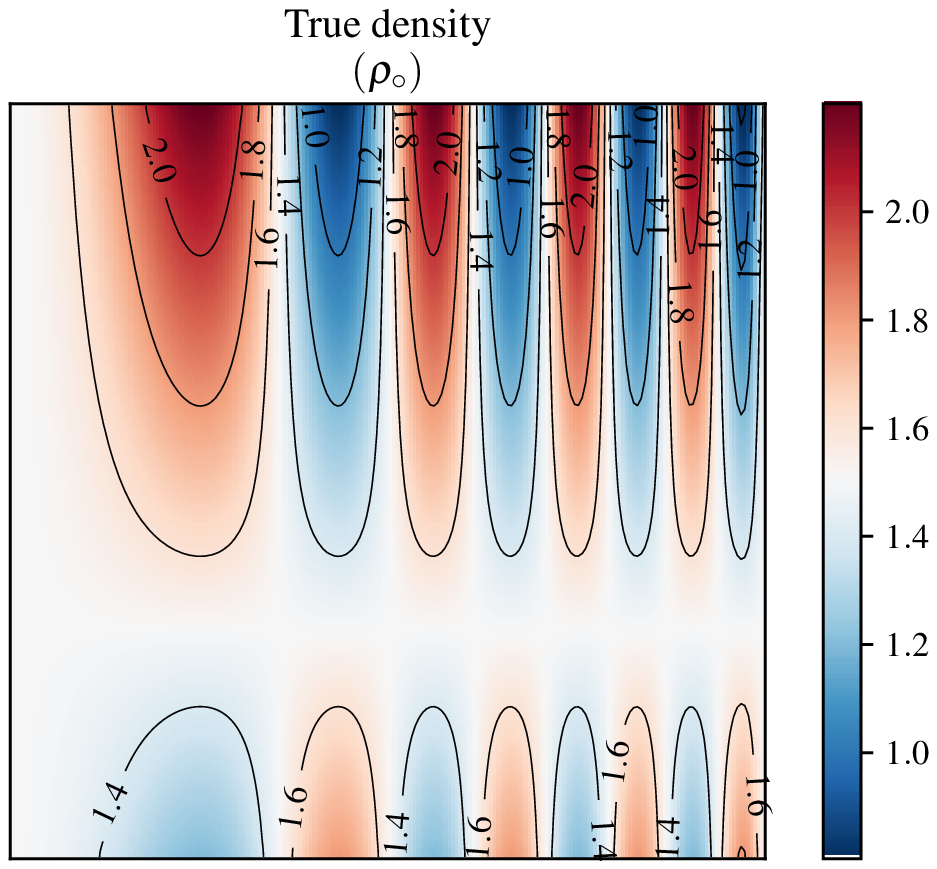}}
			\subfigure[]{\includegraphics[width=0.325\textwidth]{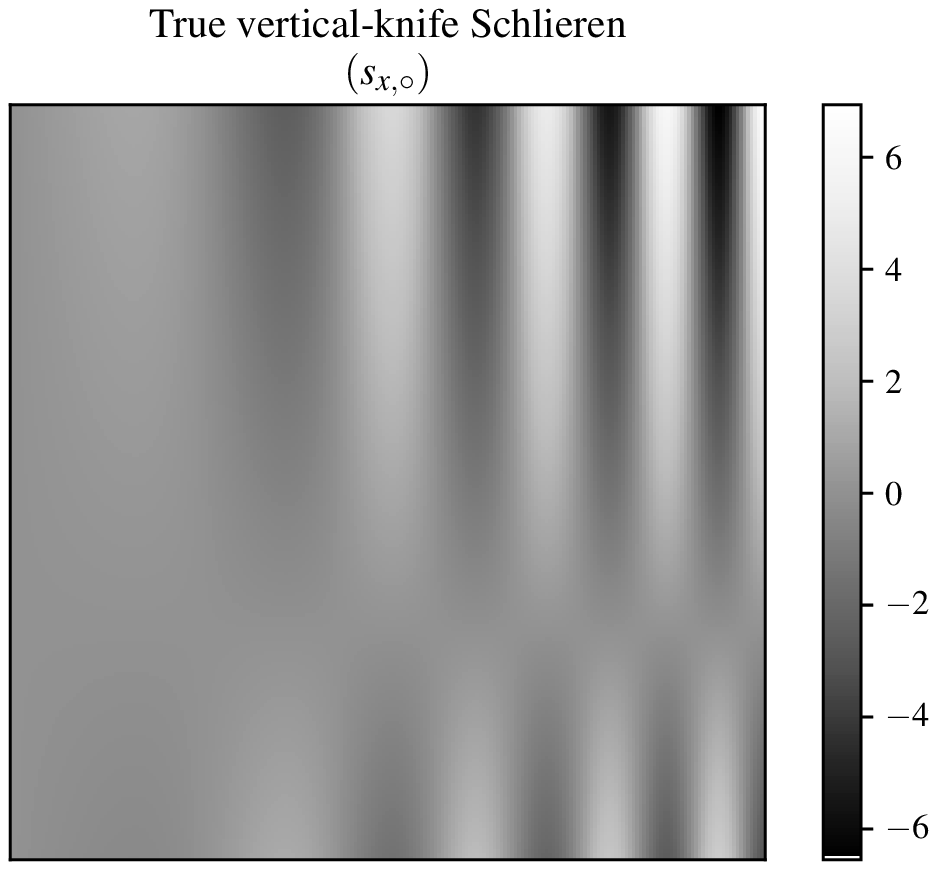}}
			\subfigure[]{\includegraphics[width=0.325\textwidth]{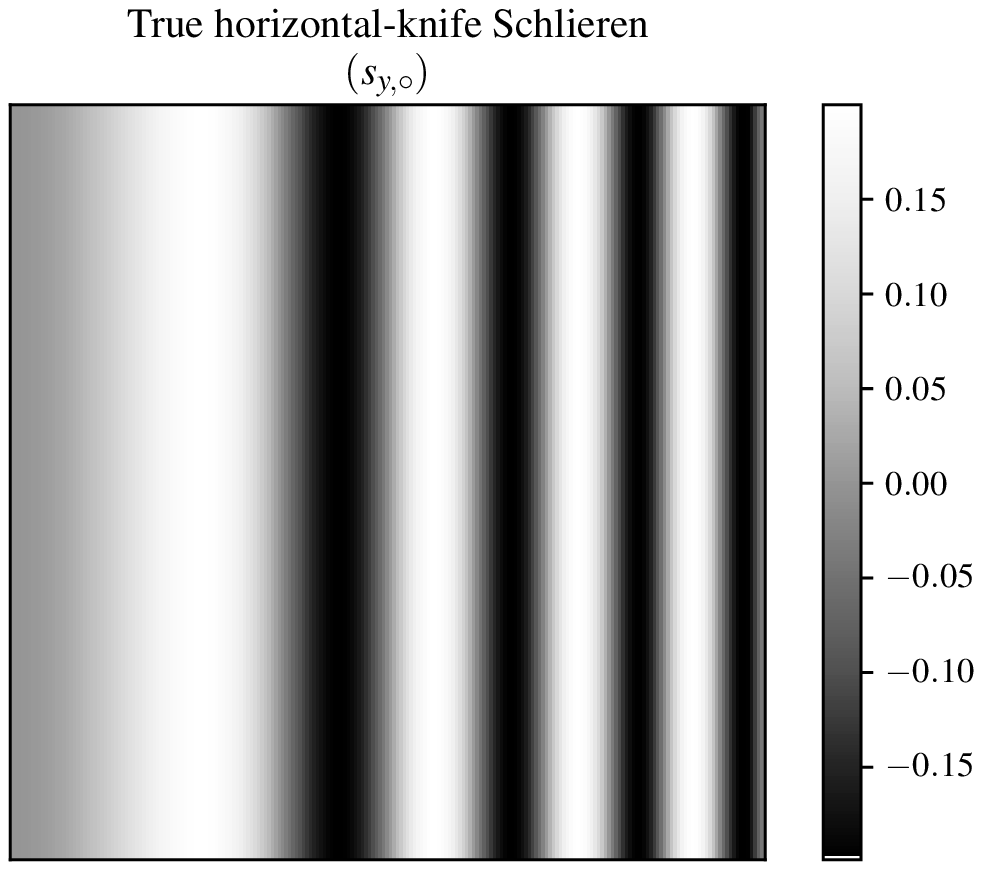}}
			\subfigure[]{\includegraphics[width=0.325\textwidth]{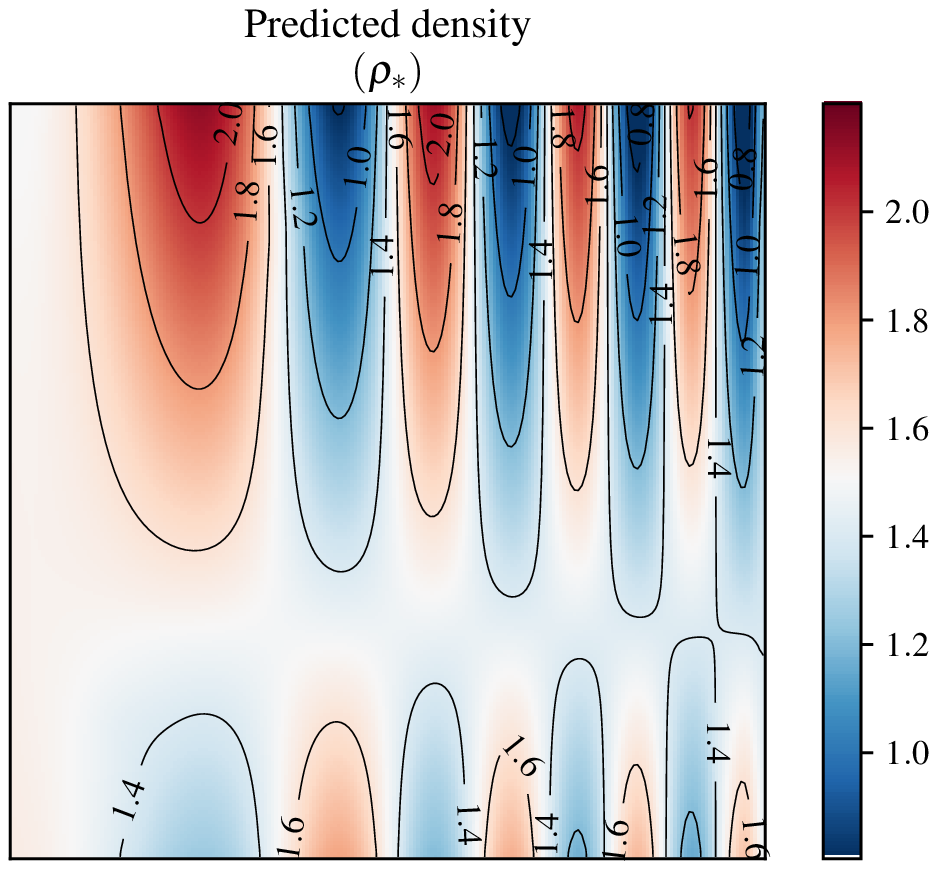}}
			\subfigure[]{\includegraphics[width=0.325\textwidth]{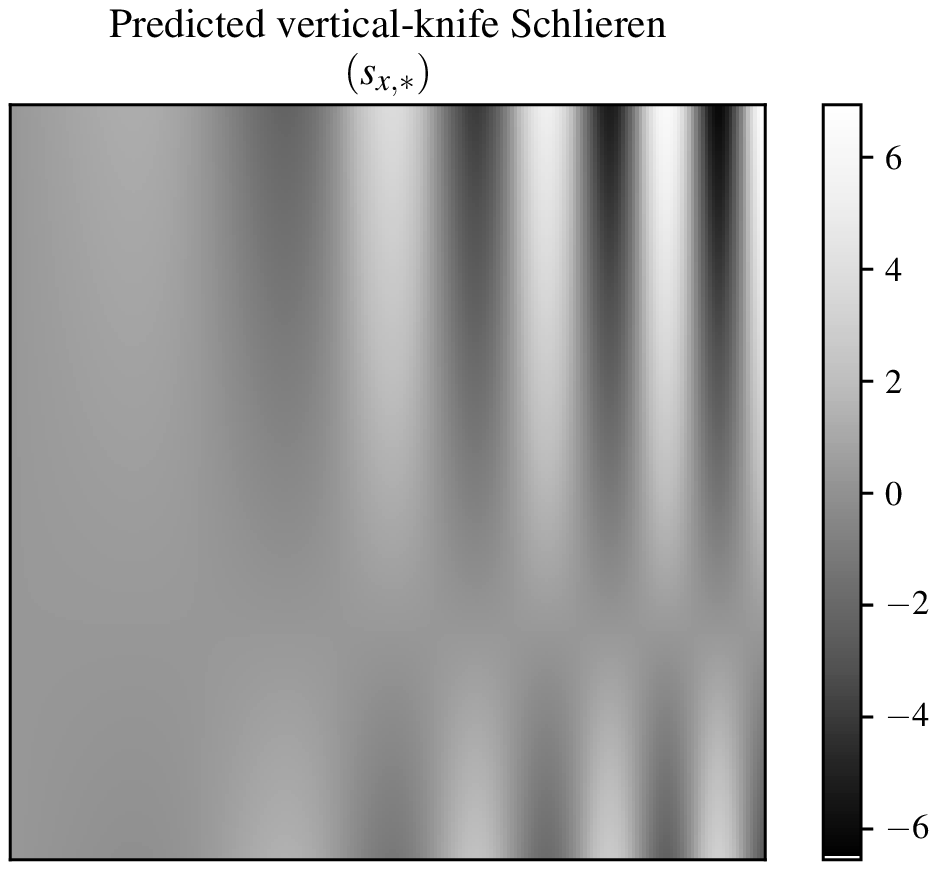}}
			\subfigure[]{\includegraphics[width=0.325\textwidth]{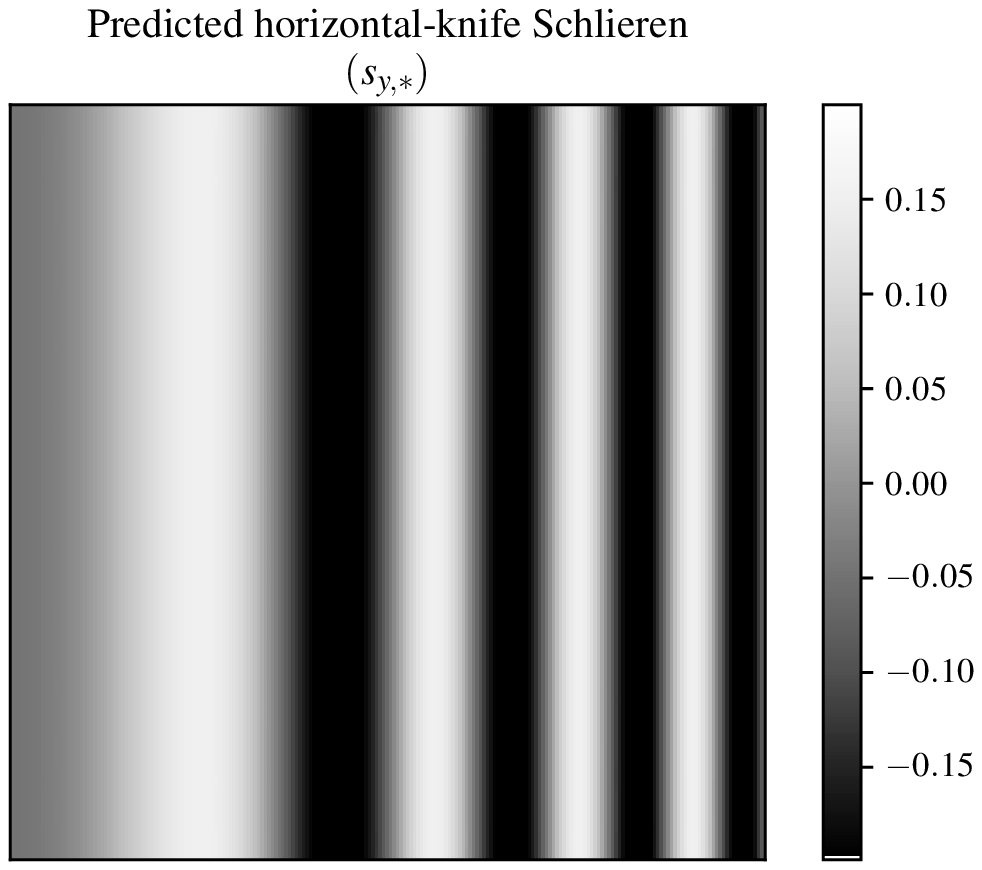}}
			\subfigure[]{\includegraphics[width=0.325\textwidth]{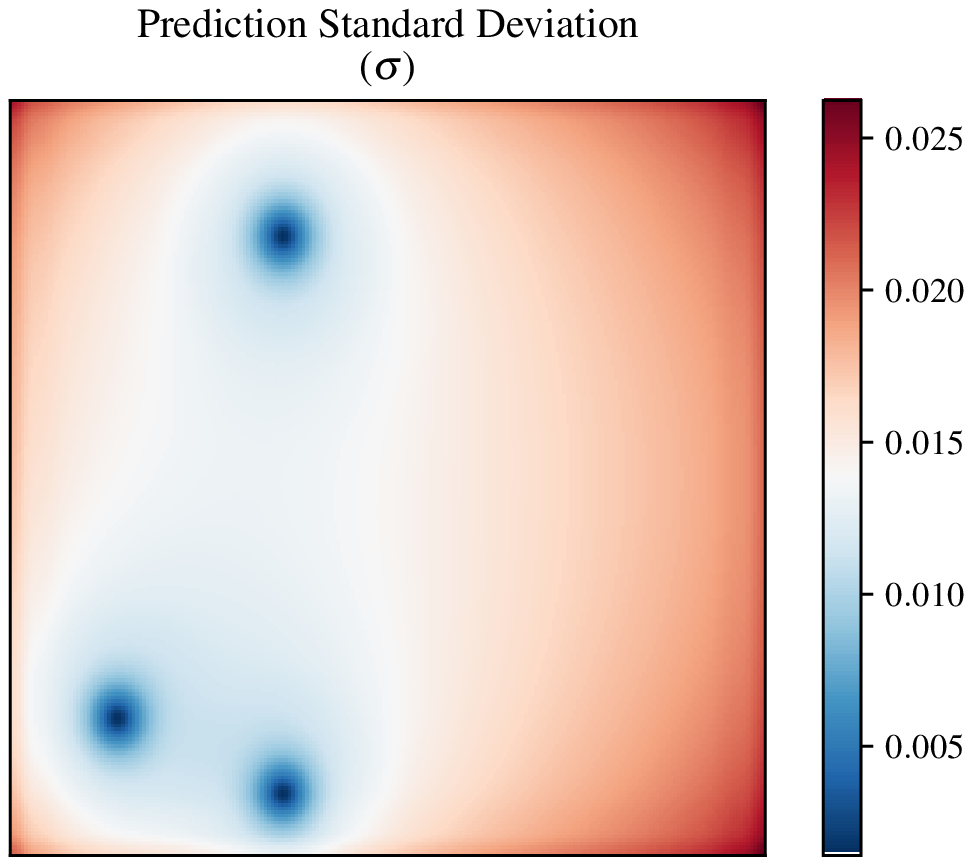}}
		\end{subfigmatrix}
		\caption{Analytical test case: (a) true density ($\rho$); (b) true vertical knife edge Schlieren image ($s_{x,\circ}$); (c) true horizontal knife edge Schlieren image ($s_{y,\circ}$); (d) GP predicted density ($\rho_{\ast}$); (e) GP predicted vertical knife edge Schlieren image ($s_{x,\ast}$); (f) GP predicted horizontal knife edge Schlieren image ($s_{y,\ast}$); (g) posterior standard deviation}
		\label{fig:analytic_results}
	\end{center}
\end{figure}

The prior definitions for this case are listed in Table \ref{tab:case1_params}, with Figure \ref{fig:analytic_results} showing a comparison of the true density (a) against the posterior predicted mean density (d), and the corresponding prediction of the gradients (e-f), and the posterior standard deviation shown in (g) exhibiting a generally uniform and small standard deviation demonstrates a high level of model confidence.

\begin{figure}[ht!]
	\begin{center}
		\begin{subfigmatrix}{2}
			\subfigure[]{\includegraphics[width=0.4\textwidth]{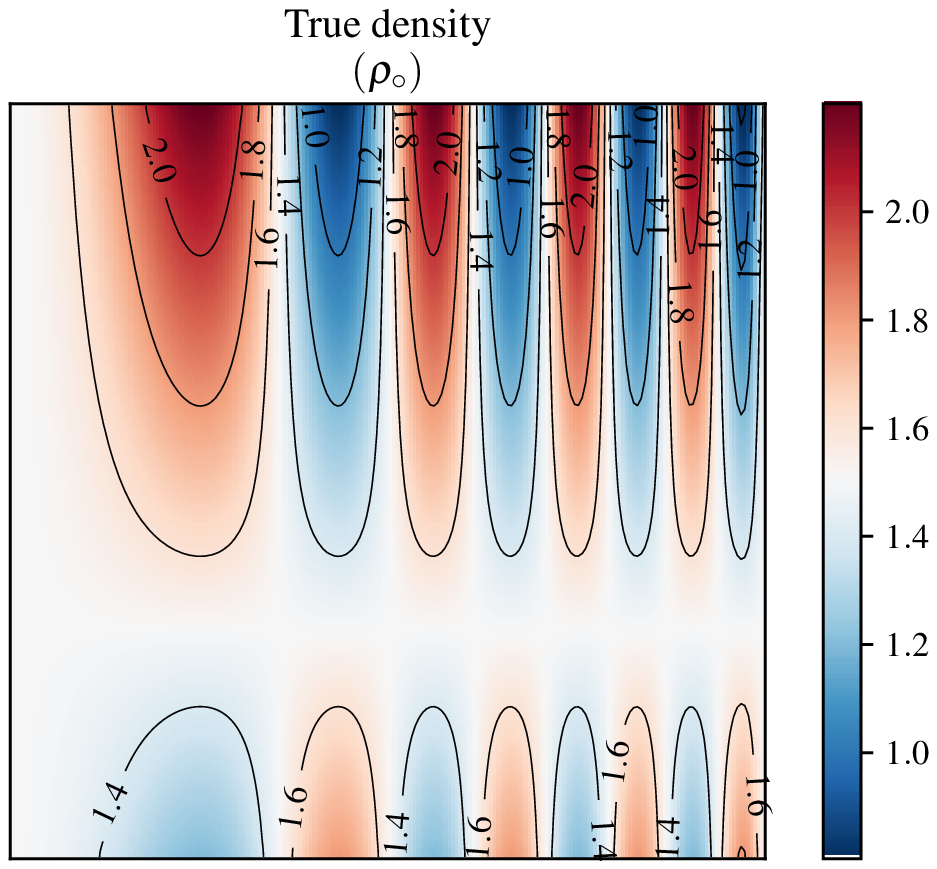}}
			\subfigure[]{\includegraphics[width=0.4\textwidth]{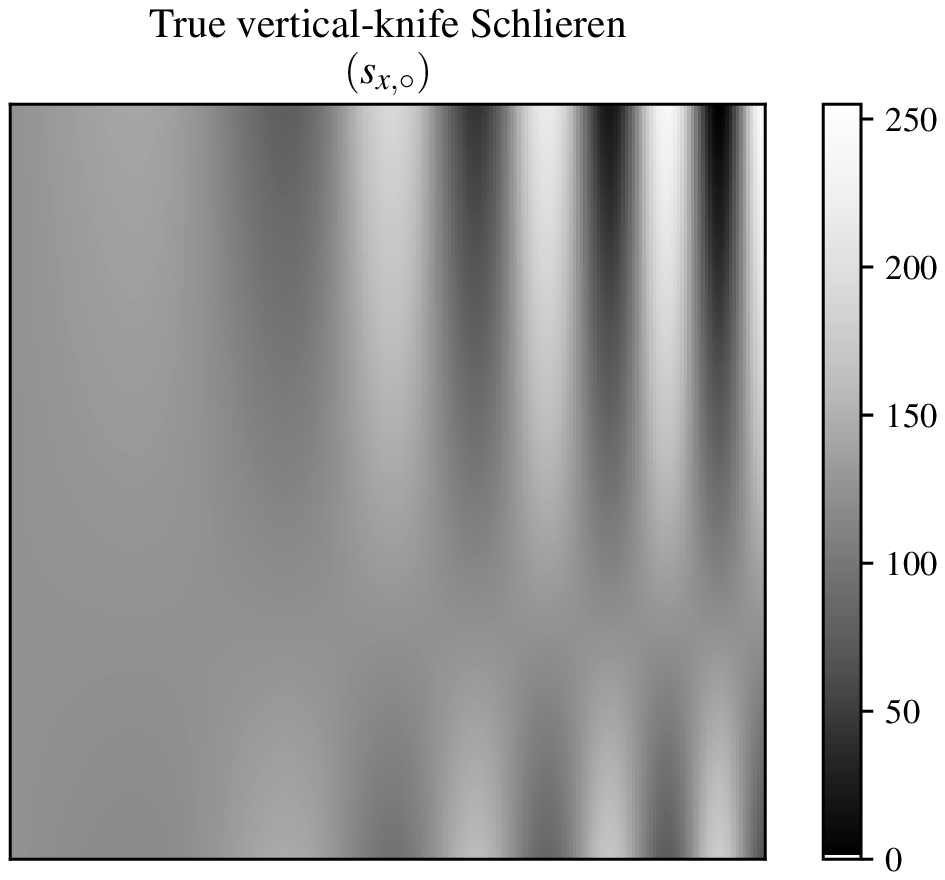}}
			\subfigure[]{\includegraphics[width=0.4\textwidth]{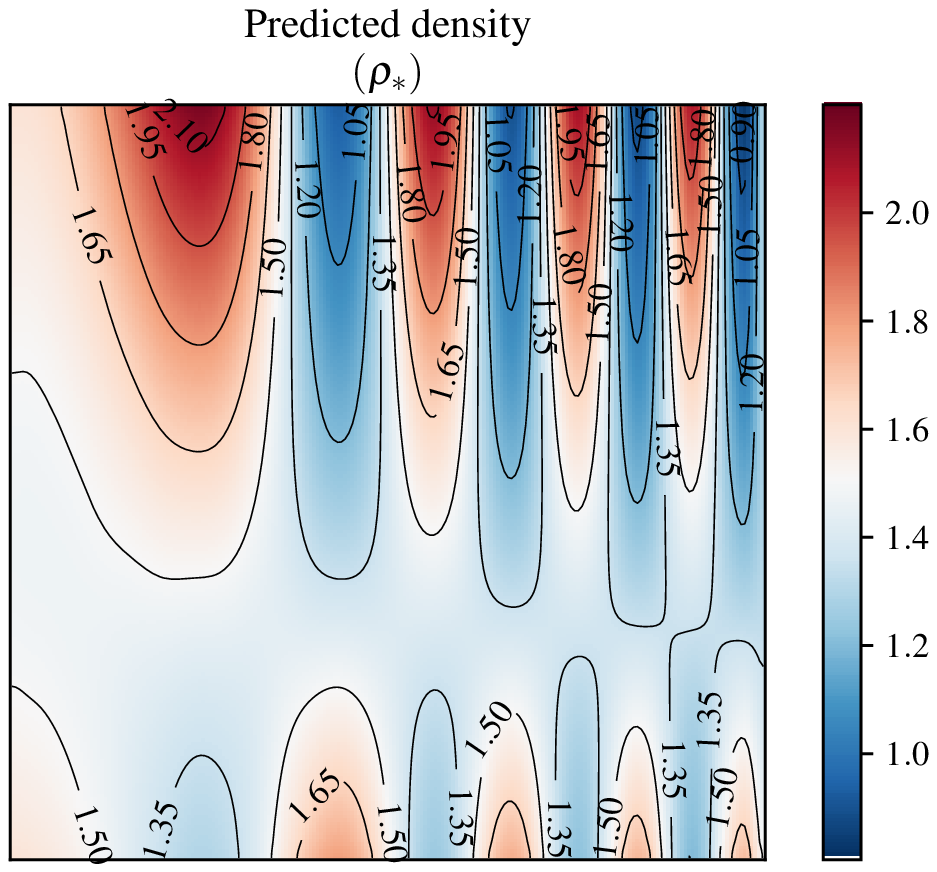}}
			\subfigure[]{\includegraphics[width=0.4\textwidth]{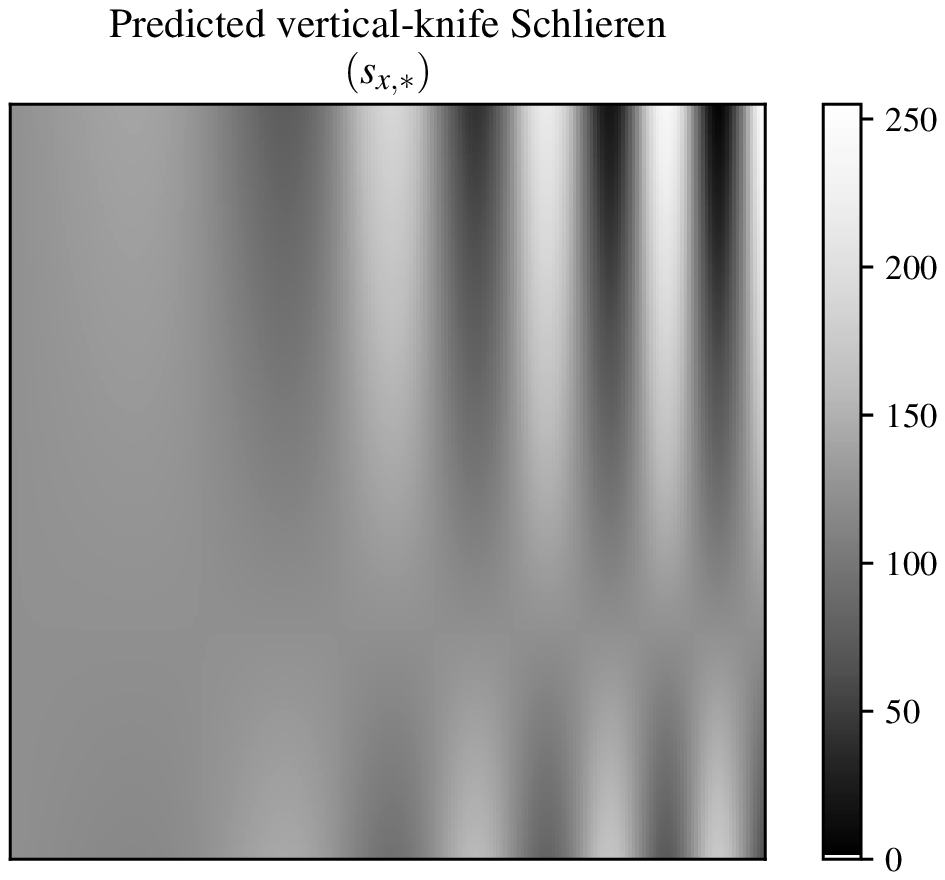}}
			\subfigure[]{\includegraphics[width=0.4\textwidth]{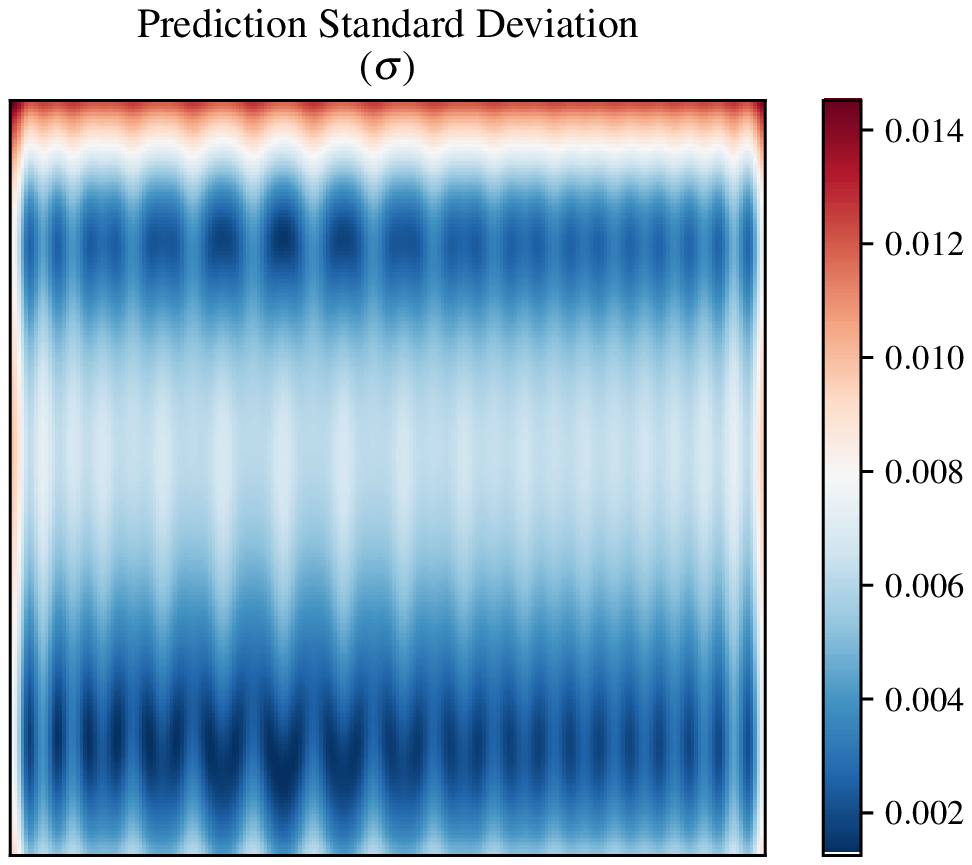}}
		\end{subfigmatrix}
		\caption{Analytical test case: (a) true density ($\rho$); (b) true vertical knife edge Schlieren image ($s_{x,\circ}$); (c) GP predicted density ($\rho_{\ast}$); (d) GP predicted vertical knife edge Schlieren image ($s_{x,\ast}$); (d) posterior standard deviation}
		\label{fig:analytic_dx_results}
	\end{center}
\end{figure}

If the flow is highly directional, and the gradients in one direction dominate the flow, such is the case in this analytical function, where the gradients in $x$ are dominant. The joint distribution can be simplified to accept gradient input from only one direction, with simpler prior definitions as listed in Table \ref{tab:case1_params_partial}. A comparison between the true and partially reconstructed density field shown in Figure \ref{fig:analytic_dx_results} shows good agreement. Since the density gradient is highly directional, this method is still, able to capture the correct trends and provide an indicative result.

\begin{table}[ht!]
	\centering
	\caption{Prior definition for case 1 with partial gradient observation}
	\label{tab:case1_params_partial}
	\begin{tabular}{@{}llll@{}}
		\toprule
		Parameter & Distribution & Moments &  \\ \midrule
		$l_x, l_y$        &  half-normal  &     $\sigma=1.0$    &  \\
		$\alpha_x, \alpha_y$    & half-normal  &    $\sigma=1.0$     &  \\
		$\xi$      & half-normal  &    $\sigma=2.0$     &  \\
		$\sigma_y$    & deterministic  &    1E-3     &  \\
		$\sigma_{dy}$    & deterministic  &    0.1     &  \\ \bottomrule
	\end{tabular}
\end{table}

\FloatBarrier
\subsection{Priors for the wind tunnel sting model}
Priors for the wind tunnel sting model Gaussian process are given in Tables \ref{tab:case2_params} and \ref{tab:case2_dx_params} for results with both schlieren images, and with only the vertical knife-edge image as inputs respectively. The lengthscales ($l_x, l_y$) for this case are defined based on the spacing between the training locations.

\begin{table}[ht!]
	\centering
	\caption{Prior definition for case 2}
	\label{tab:case2_params}
	\begin{tabular}{@{}llll@{}}
		\toprule
		Parameter & Distribution & Moments &  \\ \midrule
		$l_x$        & deterministic    &  0.67        &  \\
		$l_y$        & deterministic    &  0.675        &  \\
		$\lambda_x, \lambda_y$    & half-normal  &    $\sigma=0.5$     &  \\
		$\xi$      & half-normal  &    $\sigma=0.5$     &  \\
		$\sigma_{y}$    & deterministic  &    1E-4     &  \\
		$\sigma_{dy}$    & deterministic  &    0.1     &  \\ \bottomrule
	\end{tabular}
\end{table}

\begin{table}[ht!]
	\centering
	\caption{Prior definition for case 2 with partial gradient observation}
	\label{tab:case2_dx_params}
	\begin{tabular}{@{}llll@{}}
		\toprule
		Parameter & Distribution & Moments &  \\ \midrule
		$l_x$        & deterministic    &  0.5        &  \\
		$l_y$        & deterministic    &  0.2        &  \\
		$\lambda_x, \lambda_y$    & half-normal  &    $\sigma=0.5$     &  \\
		$\xi$      & half-normal  &    $\sigma=0.5$     &  \\
		$\sigma_{y}$    & deterministic  &    1E-2     &  \\
		$\sigma_{dy}$    & deterministic  &    0.1     &  \\ \bottomrule
	\end{tabular}
\end{table}

\FloatBarrier
\subsection{Priors for the supersonic aircraft in flight}
Priors for the wind tunnel sting model Gaussian process are given in Table~\ref{tab:case3_params}; flow properties used to estimate the density are provided in Table~\ref{tab:case3_flowcons}.

\begin{table}[ht!]
	\centering
	\caption{Prior definition for case 3}
	\label{tab:case3_params}
	\begin{tabular}{@{}llll@{}}
		\toprule
		Parameter & Distribution & Moments &  \\ \midrule
		$l_x, l_y$        &  half-normal  &     $\sigma=1.0$    &  \\
		$\lambda_x, \lambda_y$    & half-normal  &    $\sigma=1.0$     &  \\
		$\xi$      & half-normal  &    $\sigma=1.0$     &  \\
		$\sigma_y$    & half-normal  &    1.0     &  \\
		$\sigma_{dy}$    & half-normal  &    0.3     &  \\ \bottomrule
	\end{tabular}
\end{table}

\begin{table}[ht!]
	\centering
	\caption{Flow properties for case 3}
	\label{tab:case3_flowcons}
	\begin{tabular}{@{}ll@{}}
		\toprule
		Flow property & Value   \\ \midrule
		$\rho_{1}$    & 0.459 $kg/m^3$ $\pm$ 0.005 $kg/m^3$    \\
		$\rho_{2} (estimated)$    & 0.5 $kg/m^3$ $\pm$ 0.06 $kg/m^3$    \\
		$\gamma$    & 1.401 $\pm$ 0.001     \\
		$M_{1}$        &  1.05 $\pm$ 0.05 \\
		$\beta$      & 77$^{\circ}$ $\pm$ 2$^{\circ}$      \\ \bottomrule
	\end{tabular}
\end{table}

\FloatBarrier
\subsection{Priors for the shocktube CFD case}
Priors for the supersonic CFD Gaussian process are given in Table~\ref{tab:case4_params}.

\begin{table}[ht!]
	\centering
	\caption{Prior definition for case 4}
	\label{tab:case4_params}
	\begin{tabular}{@{}llll@{}}
		\toprule
		Parameter & Distribution & Moments &  \\ \midrule
		$l_x, l_y$        &  half-normal  &     $\sigma=1.0$    &  \\
		$\lambda_x, \lambda_y$    & half-normal  &    $\sigma=5.0$     &  \\
		$\xi$      & half-normal  &    $\sigma=5.0$     &  \\
		$\sigma_y$    & deterministic  &    2.5E-4     &  \\
		$\sigma_{dy}$    & deterministic  &    1.0     &  \\ \bottomrule
	\end{tabular}
\end{table}
\FloatBarrier

\paragraph{Acknowledgments}
Special thanks to Masanori Ota from Chiba University for kindly providing schlieren images of the asymmetric sting model for use as a test case. Thanks also to Tarik Dzanic from Texas A\&M University for providing the CFD results for the Sod shock tube test case. Please note that the Schlieren aircraft images in Figures~\ref{fig:process_chart} and \ref{fig:nasa_aircraft} are sourced from a United States government work \citep{Heineck2016, Heineck2021} and are therefore not subject to copyright protection. 

\paragraph{Funding Statement}
The authors were partly supported by the Lloyd’s Register Foundation-Alan Turing Institute Strategic Priorities Fund. The fund is delivered by UK Research and Innovation, with this award managed by EPSRC (EP/T001569/1). PS also wishes to acknowledge the funding and support of Rolls-Royce plc.

\paragraph{Competing Interests}
The authors declare no competing interests exist.

\paragraph{Data Availability Statement}
The data and all aspects of the code that support the findings of this study, including the figure generation are openly available in the `dce-schlieren-density-reconstruction' repository at `github.com' (https://github.com/bnubald/dce-schlieren-density-reconstruction).

\paragraph{Ethical Standards}
The research meets all ethical guidelines, including adherence to the legal requirements of the study country.

\paragraph{Author Contributions}

Conceptualization: B.N.U.; P.S. Data curation: B.N.U. Formal analysis: B.N.U.; P.S. Investigation: B.N.U.; P.S. Methodology: B.N.U.; P.S. Project administration: B.N.U.; P.S. Software: B.N.U.; P.S. Supervision: P.S.; A.D. Validation: B.N.U. Visualization: B.N.U. Writing - original draft: B.N.U. Writing - review \& editing: B.N.U.; P.S.; A.D. All authors approved the final submitted draft.

\bibliographystyle{apalike}
\bibliography{arxiv}

\end{document}